\documentclass{aa}

\usepackage{graphicx}
\usepackage{float}
\usepackage{sidecap} 
\usepackage{txfonts}
\usepackage{color}

\usepackage{pifont}

\usepackage{enumitem}
\usepackage{ulem}

\usepackage{color}

\usepackage{tikz}

\usepackage[colorlinks=true,allcolors=blue]{hyperref}

\newcommand{\NOH}{$N($OH$)$}

\newcommand{\TIR}{$T_{\rm{IR}}$}
\newcommand{\TK}{$T_{\rm{K}}$}
\newcommand{\nH}{$n_{\rm{H}}$}

\newcommand{\cmsq}{cm$^{-3}$}

\newcommand{\BT}[1]{\textcolor{black}{#1}}

\newcommand{\rev}[1]{\textcolor{black}{#1}}
\newcommand{\revbis}[1]{\textcolor{black}{#1}}
\newcommand{\LE}[1]{\textcolor{black}{#1}}

\begin{document} 

   \title{OH mid-infrared emission as a diagnostic of H$_2$O UV photodissociation}
  \subtitle{I. Model and application to the HH 211 shock}
   \author{Benoît Tabone\inst{1},
          Marc C. van Hemert\inst{2}, Ewine F. van Dishoeck\inst{1, 3}, John H. Black\inst{4}
          }
   \institute{Leiden Observatory, Leiden University, PO Box 9513, 2300 RA Leiden, The Netherlands
      \and
   Leiden Institute of Chemistry, Gorlaeus Laboratories, Leiden University, Einsteinweg 55, 2333 CC Leiden, The Netherlands 
   \and
             Max-Planck-Institut für Extraterrestrische Physik, Giessenbachstrasse1, 85748 Garching, Germany
    \and
    Department of Space, Earth and Environment, Chalmers University of Technology, Onsala Space Observatory, 43992, Onsala, Sweden        
            }    
    
\date{\today}
\abstract
{Water is an important molecule in interstellar and circumstellar environments. Previous observations of mid-infrared (IR) rotational lines of OH toward star-forming regions suggest that OH emission may be used to probe the photodissociation of water.}
{Our goal is to propose a method to \rev{quantify} H$_2$O photodissociation and measure the local ultraviolet (UV) flux from observations of mid-infrared OH lines.}
{Cross sections for the photodissociation of H$_2$O resolving individual electronic, vibrational, and rotational states of the OH fragment are collected. 
The state distribution of nascent OH following H$_2$O photodissociation is computed for various astrophysically relevant UV radiation fields (e.g., a single Ly$\alpha$ line or a \LE{broadband} spectrum).  
These distributions are incorporated in a new molecular excitation code called {\tt GROSBETA}, which includes radiative pumping, collisional (de)excitation, and prompt emission (i.e., following the production of OH in excited states). The influence of the photodissociation rate of H$_2$O, the spectral shape of the UV radiation field, the density, the temperature of the gas, and the strength of the IR background radiation field on the integrated line intensities are studied in detail. As a test case, our model is compared to \textit{Spitzer}-IRS observations at the tip of the HH~211 bow-shock.}
{The OH rotational line intensities in the range $9-16~\mu$m, covering rotational transitions with $N_{\text{\text{up}}}=18$ to $45$, are proportional to the column density of H$_2$O photodissociated per second by photons in the range $114-143~$nm (denoted as $\Phi^{\tilde{B}}$) and do not depend on other local properties such as the IR radiation field, the density, or the kinetic temperature. Provided an independent measurement of the column density of water is available, the strength of the local UV radiation field can be deduced with good accuracy, regardless of the exact shape of the UV field. 
In contrast, OH lines at longer far-IR wavelengths are primarily produced by IR radiative pumping and collisions, depending on the chemical pumping rate defined as $\mathcal{D}^{\tilde{B}} = \Phi^{\tilde{B}}/N($OH$)$ and on the local physical conditions ($n_{\rm{H}}$, $T_{\rm{K}}$, IR radiation field). Our model successfully reproduces the OH mid-IR lines in the $10-16~\mu$m range observed toward the tip of the HH 211 bow-shock and \rev{shows that the jet shock irradiates its surroundings, exposing H$_2$O to a UV photon flux \LE{that is about} $5 \times 10^3$ times larger than the standard interstellar radiation field.} We also find that chemical pumping by the reaction H$_2$ + O may supplement the excitation of lines in the range $16-30~\mu$m, suggesting that these lines could also be used to measure the two-body formation rates of OH.}
{The mid-IR lines of OH \BT{constitute a powerful diagnostic} for inferring the photodissociation rate of water and thus the UV field that water is exposed to. Future JWST-MIRI observations will be able to map the photodestruction rate of H$_2$O in various dense (\nH$\gtrsim 10^6$~cm$^{-3}$) and irradiated environments and provide robust estimates of the local UV radiation field.}           


\maketitle
   

\section{Introduction}
Oxygen is the third-most abundant element in the interstellar medium (ISM) after hydrogen and helium. 
Water is an important oxygen-bearing molecule in the formation of stars and planets.
Water ice promotes the coagulation of dust grains, from dense clouds to planet-forming disks \citep{1993ApJ...407..806C,2014prpl.conf..339T,2017A&A...602A..21S}, \LE{and water vapor is an important coolant of the gas in warm molecular environments \citep{1995ApJS..100..132N}, particularly in embedded protostellar
systems} \citep{2013A&A...552A.141K}. Ultimately, the delivery of water to young planets and small bodies conditions the emergence of life as we know it \citep{2005ARA&A..43...31C}. Following the trail of water and its associated chemistry from clouds to young forming planets is a major goal of astrochemistry \citep{2014prpl.conf..835V}. 


\LE{Infrared (IR) observations}, supplemented by chemical modeling, have demonstrated that abundant ice is built at the surface of grains in cold dense clouds, locking in $\sim 25\%$ of the oxygen \citep{2007ApJ...668..294C,2015ARA&A..53..541B}. En route to the protostellar disk, condensed H$_2$O is released into the gas phase by passive heating from the accreting protostar \citep{1996ApJ...471..400C} and jet-driven shocks \citep[e.g.,][]{2012MNRAS.421.2786F}. Under \rev{these} dense (\nH $\gtrsim 10^6$~cm$^{-3}$) and warm ($T \gtrsim 200$~K) conditions, oxygen chemistry is then driven by fast neutral-neutral reactions and photodissociation. In this context, the hydroxyl radical (OH) is a key reactive intermediate between atomic oxygen and water. The available gas-phase atomic oxygen is converted into OH by the reaction O + H$_2 \rightarrow$ OH + H, followed by the formation of water through OH + H$_2 \rightarrow$ H$_2$O + H. \rev{OH is destroyed by the latter reaction, by the former reaction in the backward direction, and by UV photodissociation.} Water is also destroyed either by the backward reaction H$_2$O + H $ \rightarrow$ OH + H$_2$ or by photodissociation leading mostly to OH. 

The observations of warm H$_2$O and OH vapor toward embedded protostars culminated with the \textit{Herschel Space Observatory} \citep{2011PASP..123..138V}, which outperformed previous far-IR and submillimeter observatories (e.g., SWAS, Odin, ISO-LWS). The detection of far-IR OH lines in inner envelopes and outflows \citep{2010A&A...521L..36W,2011A&A...531L..16W,2013A&A...552A..56W,2015ApJ...799..102G}, \rev{supplemented by similar detections in prototypical \LE{photodissociation} regions \citep{2011A&A...530L..16G,2017A&A...599A..20P} and extragalactic sources \citep{2011ApJ...733L..16S,2012A&A...541A...4G}} evidenced the presence of an active warm oxygen chemistry in dense interstellar environments. One of the most striking results is the relatively low abundance of gaseous water found toward most of the \rev{protostellar sources}, which contrasts with the abundant water ice supplied by the infalling envelope. In particular, multiple outflow components, which dominate H$_2$O emission, exhibit abundances ranging from $10^{-7}$ to $10^{-5}$ \citep{2012A&A...542A...8K,2013A&A...549A..16N,2017A&A...605A..93K}. Water vapor from envelopes and perhaps embedded disks is typically traced by H$_2^{17}$O and H$_2^{18}$O lines and is best observed by \textit{Herschel}-HIFI and (sub)millimeter interferometers (NOEMA, SMA, ALMA). While \citet{2013ApJ...769...19V} show that water abundance can be as high as $10^{-4}$, most of the protostellar systems exhibit lower water abundances, typically ranging between $10^{-7}$ and $10^{-5}$ \citep{2010ApJ...710L..72J,2012A&A...541A..39P,2016A&A...590A..33P,2020A&A...636A..26H}. Following the dissipation of the envelope, \LE{near- and mid-IR} observations from space (e.g., \textit{Spitzer}) and from the ground (e.g., VLT-CRIRES) \LE{have unveiled abundant hot water vapor in the planet-forming regions of Class II disks \citep[$\lesssim 10~$au,][]{2014prpl.conf..363P} but surprisingly dry Herbig disks \citep{2010ApJ...720..887P,2011ApJ...732..106F,2015A&A...582A..88W}. }

The solution to this conundrum of low water abundance may lie in the impact of the ultraviolet (UV) radiation field produced by the accreting young star or strong jet shocks ($V \gtrsim 40~$km~s$^{-1}$). Observations of hydrides provide evidence of the driving role of UV irradiation in the chemistry of protostars \citep{2016A&A...590A.105B}. It has been proposed that the low abundance of H$_2$O in protostellar outflows is the result of the UV photodissociation of H$_2$O vapor \citep{2002ApJ...574..246N,2013A&A...549A..16N,2014A&A...572A...9K}. The low water abundance in embedded protostellar disks is generally attributed to the freeze-out of water \citep{2009A&A...495..881V}. However, recent constraints on the thermal structures of Class 0 disk-like structures may contradict this interpretation \citep{2020A&A...633A...7V} and require the efficient destruction of H$_2$O or a large opacity of the dust in the far-IR. 

The impact of UV photodissociation is generally explored by means of astrochemical models \citep{2010A&A...518L.121V,2012A&A...537A..55V,2012A&A...538A...2P,2016A&A...585A..74Y}. However, the strength of the local UV radiation field incident on disks, envelopes, and outflows is poorly known, due to the complex sources of UV emission and the unknown level of dust and gas attenuation on the line of sight, severely limiting the diagnostic capabilities. Assessing the exact role of UV photodissociation of water from astrochemical models based only on molecular abundances remains highly uncertain.

Alternatively, the excitation of molecular species provides in some cases a direct access to their formation and destruction routes. In irradiated regions, IR lines of H$_2$ and CO from ro-vibrationally excited levels trace the UV pumping followed by radiative decay \citep{1980ApJ...240..940K,1987ApJ...322..412B} and have been used to probe the photodissociation of these species \citep{2007A&A...467..187R,2013A&A...551A..49T}. The excitation of reactive species for which the collisional time scale is comparable to the destruction time scale, carry also key information on their formation and destruction rates through a process called \LE{''chemical pumping''} \citep{1998FaDi..109..257B}. So far, the incorporation of a state-to-state chemistry in models (i.e., models considering the state distribution of the reagents and the products) have been carried out only for a handful of molecular ions relevant for diffuse ISM and early universe \citep[][]{2013A&A...550A...8G,2009A&A...505..195S,2013MNRAS.434..114C}.

Mid-infrared observations with \textit{Spitzer}-IRS toward protostellar outflows and young disks evidenced the presence of a superthermal population of rotationally excited OH with energies up to $E_{\text{up}} \simeq 28 000$~K \citep{2008ApJ...680L.117T,2012ApJ...751....9T,2014ApJ...788...66C}.
\LE{Two decades of extensive quantum calculations  supplemented by laboratory experiments have permitted the direct interpretation of these observations as the smoking gun of H$_2$O photodissociation.} Specifically, seminal works have shown that H$_2$O photodissociation through the $\tilde{A}$ excited electronic state of H$_2$O ($ \lambda \gtrsim 143$~nm) produces OH in vibrationally hot but rotationally cold states \citep[e.g.,][]{1999JChPh.110.4119H,2000JChPh.11310597Y,2001JChPh.114.9453V}, whereas photodissociation through the $\tilde{B}$ state ($ 114 \lesssim \lambda \lesssim 143$~nm) produces OH in high rotational states with levels up to $N \simeq 47$ \citep[$E\simeq 45 000$ K,][]{2000JChPh.11310073H,2000JChPh.112.5787V}. \LE{The rotationally excited lines of OH would thus probe the radiative cascade of OH products following their formation via H$_2$O photodissociation, a process called ''prompt emission''}. If consistently modeled, rotational lines of OH can thus give a direct access to the photodissociation of H$_2$O. However, a detailed modeling that includes chemical data accumulated over the past decades is lacking.

Probing H$_2$O photodissocation in astrophysical environments may also lead to unique constraints on the local UV irradiation field that H$_2$O is exposed to. The \LE{far-UV} radiation field is a key parameter that controls the chemical, physical, and dynamical evolution of circumstellar regions. It regulates the overall chemistry of disk upper layers \citep[e.g.,][]{2006FaDi..133..231V,2012ApJ...747..114W} and outflows \citep{2012A&A...538A...2P,2020A&A...636A..60T}, the thermal structure of disks \citep{2004ApJ...613..424G,2012A&A...541A..91B,2016A&A...586A.103W}, and the coupling between the gas and the magnetic field \citep[e.g.,][]{1996ApJ...457..355G,2019ApJ...874...90W}. The impact of the UV radiation field depends not only on attenuation processes, but also on its spectral shape. While the interstellar radiation field exhibits a broadband emission down to 91~nm \citep{1968BAN....19..421H,1978ApJS...36..595D,1983A&A...128..212M}, the UV radiation emitted by accreting nascent stars \citep{2003ApJ...591L.159B,2004ApJ...607..369H,2012ApJ...744..121Y,2012ApJ...756L..23S} or by strong jet shocks \citep[$\gtrsim 40$~km~s,$^{-1}$][]{1979ApJS...39....1R,2017ApJS..229...35D} is often dominated by Lyman-$\alpha$ emission, leading to the selective photodissociation of species that exhibit large photodissociation cross sections at $121.6~$nm. 

To date, few robust diagnostics have been proposed to directly access the local UV radiation field. H$_2$ and CO lines excited by UV pumping probe only UV photons in narrow lines at $< 114~$nm that are rapidly attenuated and are not indicative of the broadband UV spectrum relevant for the other UV photoprocesses. Chemical diagnostics have also been proposed but they rely either on H$_2$ UV pumping \citep[e.g., CN formed from excited H$_2$,][]{2018A&A...609A..93C} or depend on elemental abundance ratios \citep[e.g., hydrocarbons,][]{2016ApJ...831..101B}. 

In this work, we explore the potential of the OH mid-IR lines for probing H$_2$O photodissociation and the local UV radiation field under a broad range of physical conditions representative of dense irradiated environments, such as molecular shocks, circumstellar media, and \rev{prototypical} \LE{photodissociation} regions (PDRs). To reach this goal, results of quantum mechanical calculations resolving the electronic, vibrational, and rotational states of the OH product following H$_2$O dissociation at different UV wavelengths are collected. The distribution of the OH fragments following H$_2$O photodissociation by UV fields of various spectral shapes are derived. The emerging line intensities are then calculated using {\tt GROSBETA}, a new molecular excitation code that includes prompt emission, radiative decay, collisional (de)excitation, and IR radiative pumping in a slab approach. For sake of conciseness, ro-vibrational lines of OH are not studied in this work.

In Sect. \ref{sec:model}, we present the OH model and the basics of {\tt GROSBETA}. The competition of the different excitation processes on the OH line intensities are studied in detail for H$_2$O photodissociation at Ly$\alpha$ ($\lambda = 121.6~$nm) and extended to other UV radiation fields in Sect. \ref{sec:results}. In Sec. \ref{sec:disscu}, a method to observationally derive the amount of H$_2$O photodissociated per unit time, the photodissociation rate of H$_2$O, and the strength of the local UV radiation field is proposed and applied to \textit{Spitzer}-IRS observations of the tip of the HH 211 protostellar bow-shock. Our findings are summarized in Sect. \ref{sec:concl}.

\section{Model}
\label{sec:model}
\subsection{OH model}

\subsubsection{Energy levels}

\begin{figure*}
\centering
\includegraphics[width=.75\textwidth]{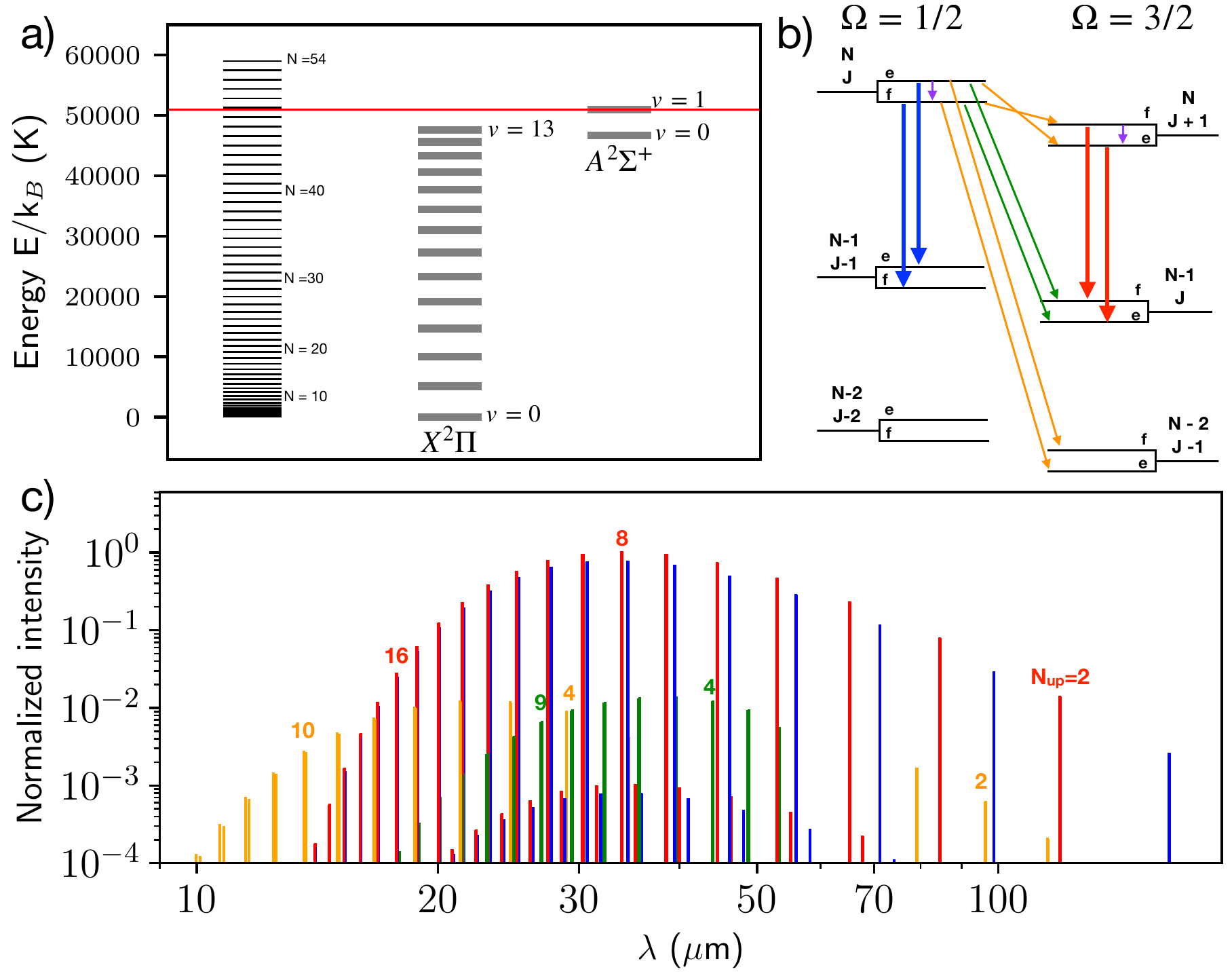}
\caption{OH model adopted in this work. a) $X^2 \Pi$ and $A^2 \Sigma^+$ electronic levels split into vibrational levels and further split into rotational levels labeled by $N$ (left ladder corresponding the OH(X$^2 \Pi$)($\varv = 0$) state). \rev{The red line indicates the dissociation energy of OH(X$^2 \Pi$).} b) Structure of the rotational ladders of OH(X$^2 \Pi$) within a vibrational state that gives rise to mid- and far-IR lines. Each $N$ level is split by the spin-orbit coupling and the $\Lambda$-doubling. The two spin-orbit states are labeled by the $\Omega$ quantum number and the $\Lambda$-doubling states are labeled by their $\epsilon=e/f$ spectroscopic parity. Radiative transitions included in our model and emerging from the four $N$-levels are also depicted by arrows. There are three kinds of transitions: intra-ladder rotational transitions (blue and red arrows), cross-ladder transitions (orange, with $\Delta N = +1$, and green, with $\Delta N = 0, 2$), and $\Lambda$-doubling transitions (purple). c) Optically thin LTE spectrum of OH at $T_K=750$~K. The color code is the same as panel b) and is repeated in Fig. \ref{fig:OH-spectra-ex} and \ref{fig:lines-grid-mid-IR}. The upper $N$ level is indicated for some transitions. The $\Lambda$-doubling lines are too weak and at longer radio wavelengths to appear in this spectrum. Ro-vibrational lines are at shorter wavelengths than shown here ($\lambda \lesssim 2.7~\mu$m).}
\label{fig:OH-rot-structure}
\end{figure*}

H$_2$O photodissociation can produce OH with high rotational and vibrational quantum numbers in the $X^2 \Pi$ ground and the $A^2\Sigma^+$ first excited electronic states. Our OH model includes the energy levels provided by \citet{2016JQSRT.168..142B} and \citet{2018JQSRT.217..416Y} and made available on the EXOMOL database\footnote{http://www.exomol.com/, \citet{2016JMoSp.327...73T}}. Figure \ref{fig:OH-rot-structure}-a shows the electronic, vibrational, and rotational levels included in the model. Regarding the ground electronic state OH($X^2 \Pi$), vibrational levels up to $\varv=13$ \LE{were included}. We retained rotational levels that are stable against dissociation and in particular included energy levels above the dissociation energy of the OH($X$) that are stabilized by the centrifugal barrier \citep{2019NatCo..10.1250C}.
This corresponds to a quantum number of $N=54$ in the $\varv=0$ state, where $N$ denotes the rotational quantum number associated with the motion of the nuclei. 
Each rotational level is further split into two spin-orbit manifolds corresponding to projected quantum numbers of the sum of the electronic orbital and spin angular momenta of $\Omega = 3/2$ and $\Omega = 1/2$ (see Fig. \ref{fig:OH-rot-structure}-b). This results in two rotational ladders with fine-structure levels characterized by a total \BT{angular momentum} quantum number of $J=N+1/2$ and $J=N-1/2$, respectively. Lastly, each rotational level ($N, \Omega$) is further split by the $\Lambda$-doubling into two levels labeled by their spectroscopic parity index\footnote{The spectroscopic parity index is related to the parity $p=+/-$ by the relation $\epsilon = p(-1)^{J-1/2}$, where $\epsilon=1$ for $e$ states and $\epsilon=-1$ for $f$ states.} $e/f$. Our model includes the fine structure and $\Lambda$-splitting of the OH($X^2 \Pi$)($\varv,N$) states. Hyperfine structure is not considered in this work.


Regarding the OH($A^2\Sigma^+$) state, levels with $\varv \ge 2$ are dissociative \citep{1983JChPh..79..873V} and \BT{the $\varv = 0$ with $N\ge 26$} and the $\varv = 1$ with $N\ge 17$ levels are predissociated \citep{1992JChPh..97.1838Y}. Therefore, \LE{we limited} the OH($A$) levels to lower $N$ of the $\varv=0,1$ states. The two rotational ladders emerging from the spin-orbit coupling of the OH($A^2\Sigma^+$)($\varv$) states \LE{were} also taken into account. In the following, the electronic, vibrational, rotational, fine-structure, and parity states of OH are labeled as $\Lambda$, $\varv$, $N$, $\Omega$, and $\epsilon$, respectively.

\subsubsection{Radiative transitions}

Our OH model includes 54276 radiative transitions with their corresponding Einstein-$A$ coefficients from \citet{2016JQSRT.168..142B} and \citet{2018JQSRT.217..416Y}. The mid- and far-IR lines of OH originate from radiative transitions occurring within vibrational states ($\Delta \varv = 0$) of the OH($X$) state (Fig. \ref{fig:OH-rot-structure}-c). Figure \ref{fig:OH-rot-structure}-b shows the radiative transitions that connect levels within a vibrational level of the $X^2 \Pi$ ground state. The radiative transitions with the highest Einstein-$A$ coefficients are the intra-ladder rotational transitions $N \rightarrow N-1$ (or equivalently $J \rightarrow J-1$) that preserve the e/f parity. These transitions give rise to lines from the mid- to the far-IR (see Fig. \ref{fig:OH-rot-structure}-c). The cross-ladder transitions connecting the $\Omega = 1/2$ and $\Omega = 3/2$ states are of two kinds: $J$-conserving ($\Delta N=-1$) and e/f conserving transitions, and $J \rightarrow J \pm 1$ ($\Delta N = 0,-2$) and e/f parity changing transitions. Einstein-$A$ coefficients of the latter two kinds are at least an order of magnitude lower than intra-ladder transitions and can be used to measure the opacity of the intra-ladder lines. Lastly, the $\Lambda$-doubling transitions, that lie from centimeter to sub-millimeter wavelengths are also considered in this work. Rovibrational transitions follow the same selection rules and any rovibrational transition accompanied by any change of vibrational quantum number $\varv' \rightarrow \varv''$ is included in this work. Rovibrational lines $\varv' = 1 \rightarrow \varv'' = 0$ typically lie at near-IR wavelengths, around 2.7~$\mu$m. 
Transitions connecting to the OH($A^2\Sigma^+$) state are electronic dipole allowed at near UV wavelengths.

\subsection{Excitation by H$_2$O photodissociation}
\label{subsec:chem-dataset}

\begin{figure}
\centering
\includegraphics[width=.47\textwidth]{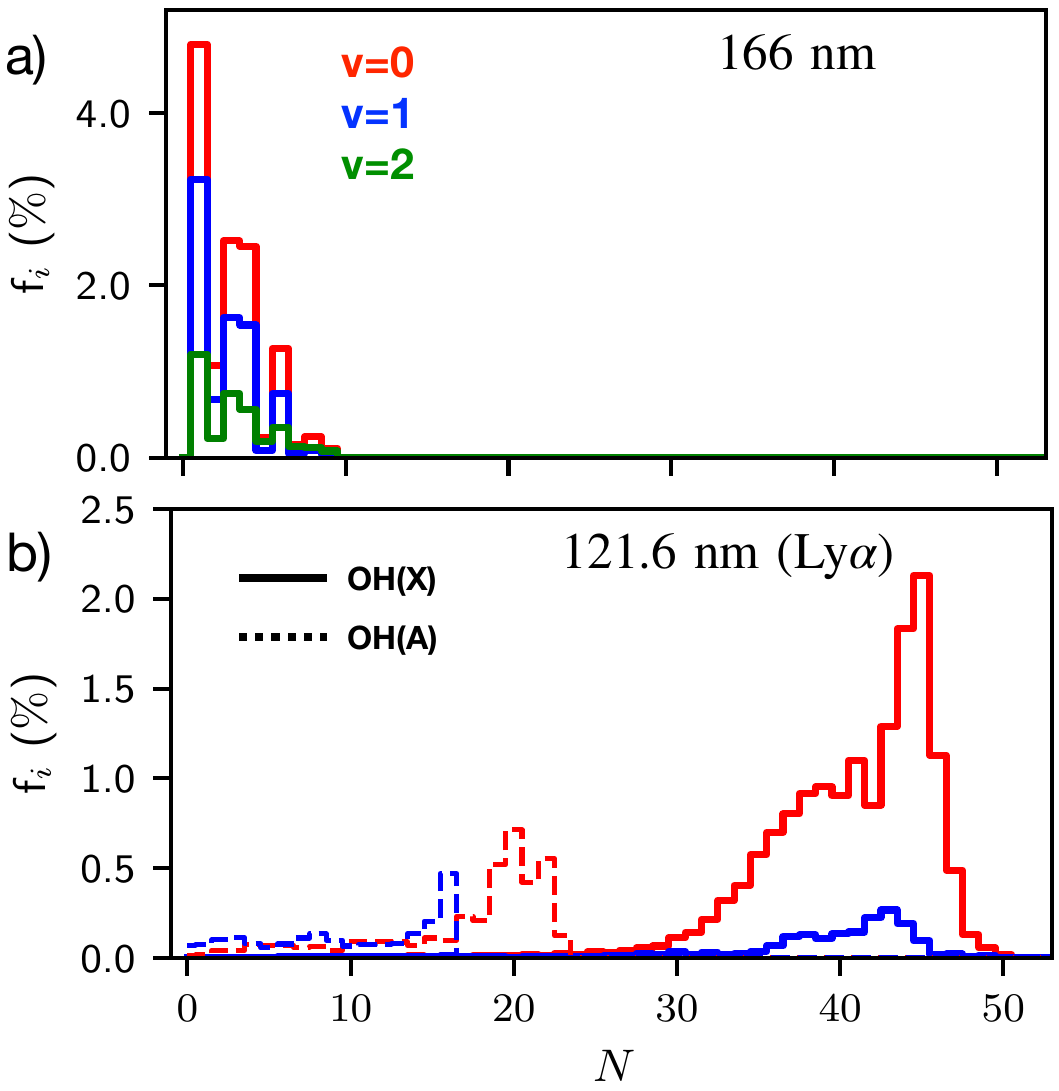} 
\caption{State distribution of the OH product following H$_2$O photodissociation at two photon wavelengths. a) \LE{At $\lambda = 166$~nm, H$_2$O dissociating via its first excited $\tilde{A}$ electronic state and producing OH in rotationally cold but vibrationally hot states.} b) \LE{At $\lambda = 121.6$~nm, H$_2$O dissociating via its second excited $\tilde{B}$ electronic state and producing OH in rotationally hot but vibrationally cold states.}
Vibrational quantum numbers are color coded as indicated in the top panel. The distributions within OH(X$^2 \Pi,\varv$) and OH(A$^2\Sigma^+,\varv$) are plotted in solid and dashed lines, respectively, as a function of the rotational quantum number $N$. Only vibrational levels that contribute at least 0.1$\%$ to the population of nascent OH are shown.}
\label{fig:example-distrib}
\end{figure}

The OH state-specific formation rate associated with the photoprocess 
\begin{equation}
\text{H}_2\text{O} + h\nu \rightarrow \rm{OH}(\lambda,\Lambda,\varv,N,\Omega, \epsilon) + \rm{H}
\label{eq:photodisso}
\end{equation}
can be expressed as
\begin{equation}
k(\Lambda, \varv,N,\Omega,  \epsilon) = \int_{\lambda} \eta(\lambda, \Lambda, \varv,N,\Omega, \epsilon) \bar{\sigma} (\lambda) I(\lambda) d\lambda
\label{eq:def-ki}
\end{equation}
\rev{and measured in cgs units in s$^{-1}$}. In this equation, $\eta(\lambda, \Lambda, \varv,N,\Omega, \epsilon)$ is the probability of forming OH in a given state following H$_2$O photodissociation by a photon of wavelength $\lambda$, $\bar{\sigma}$ is the total photodissociation cross section \rev{([cm$^2$])} of H$_2$O $\rightarrow$ OH + H at that wavelength, and $I(\lambda)$ is \BT{the photon flux averaged over all incidence angles \rev{([photon~cm$^{-2}$~s$^{-1}$~cm$^{-1}$])}.} 

\LE{We assumed} that the fine structure and $\Lambda$-doubling states of OH are equally populated by the photoprocess (\ref{eq:photodisso}) so that $\eta$ does not depend on $\Omega$ and $\epsilon$. It is also convenient to define the nascent state distribution of the OH fragments as
\begin{equation}
f(\Lambda, \varv,N) = \frac{k(\Lambda, \varv,N)}{\bar{k}},
\label{eq:def-fi}
\end{equation}
where $\bar{k}$ is the total rate for the photdissociation process (\ref{eq:photodisso}). For a monochromatic UV radiation field emitting at a wavelength $\lambda$, $f_i = \eta(\lambda,i)$. 

The absorption of a UV photon in the range $143$ to $190$~nm excites H$_2$O in its first excited singlet state $\tilde{A}$ (first absorption band) and leads to a direct dissociation to OH + H. In this work, \LE{we adopted} the OH state-specific cross sections computed by \citet{2001JChPh.114.9453V} using a wave packet approach. These data successfully reproduce the total photodissociation cross section of water in the considered energy range as well as the OH distributions measured by time-of-flight spectroscopy techniques \citep{2000JChPh.11310597Y}. Figure \ref{fig:example-distrib}-a shows the distribution of the OH fragments following  photodissociation by a photon of wavelength $\lambda = 166$~nm, where the cross section of the first absorption band peaks. Throughout the first absorption band, OH is produced in a rotationally cold ($N\lesssim 9$) but vibrationally hot state (see also Appendix \ref{app:distrib}). The proportion of vibrationally excited states increases with photon energy. 

The absorption of a UV photon in the second absorption band ($\tilde{B}$ state of H$_2$O) results in a nonadiabatic transition, leading to OH in its ground electronic state and in a direct dissociation forming electronically excited OH. The second absorption band produces a broad continuum bump in the photodissociation cross section of H$_2$O in the range $114$ to $143$~nm. However, shortward of $124.6$~nm, this dissociation channel coexists with the third and fourth absorption bands that result in sharper peaks in the photodissociation cross section (see Fig. \ref{app:cross-section-H2O}). As argued by \citet{2008JPCA..112.3002V}, the $\tilde{C}$ and $\tilde{D}$ states are bound states and predissociated by the $\tilde{B}$ state. \LE{We consequently assumed} that these channels lead to the same state distribution of the OH fragments as that via the $\tilde{B}$ state at the same photon energy. This is further supported by the agreement between experiments at $\lambda \le 124$~nm \citep[e.g., Ly$\alpha$,][]{2000JChPh.11310073H} and theoretical work considering only the fragmentation dynamics from the $\tilde{B}$ state \citep{2000JChPh.112.5787V,2008JPCA..112.3002V}. In the following, we use the term "photodissociation through the $\tilde{B}$ state" to refer to photodissociation in the range $114$ to $143~$nm. \BT{In order to obtain the relevant OH distributions, we repeated the quantum wave packet calculations as described in \citet{2000JChPh.112.5787V}, but using the Dobbyn and Knowles potential \citep{1997MolPh..91.1107D} instead of the Leiden potential since it was found that it gave better agreement with experiments \citep{2001JPCA..10511414F}.}
In this wavelength range, H$_2$O photodissociation also leads to O with a branching ratio computed by \citet{2008JPCA..112.3002V}. We used the total photodissociation cross section measured by \citet{2005CPL...416..152M} that includes features due to the $\tilde{C}$ and $\tilde{D}$ states of H$_2$O (see Fig. \ref{app:cross-section-H2O}). The photodissociation cross section of H$_2$O $\rightarrow$ OH + H was then derived by taking into account the branching ratio of the OH forming channel as described by \citet{2017A&A...602A.105H}.

As an example, we show in Fig. \ref{fig:example-distrib}-b the distibution of OH products at Ly$\alpha$ wavelength ($\lambda = 121.6$~nm). In contrast to photodissociation by lower energy photons, photodissociation in the range $114$ to $143$~nm produces OH($X^2\Pi$) with low vibrational excitation but high rotational excitation. OH is produced preferentially in the ground vibrational and electronic state, and the resulting rotational distribution within this state peaks around $N=45$, corresponding to an energy level of $\simeq 45 000$~K. 
As shown in Appendix \ref{app:distrib} (see Fig. \ref{app:eta}), the peak in the rotational distribution increases with the photon energy from $N \simeq 35$ up to $N \simeq 49$. The other vibrational states follow a similar rotational distribution as that of the $\varv= 0$ state, but their relative contribution is much smaller, with less than 5$\%$ of OH produced in each of the OH$(X)(\varv \ge 1)$ states at Ly$\alpha$. A fraction of OH is also produced in the excited electronic state OH($A$) for $\lambda<137~$nm. 
The rotational distributions of the $\varv = 0, 1$ states exhibit the same pattern as that of OH($X$) but shifted toward lower $N$ numbers. 

Shortward of $\lambda = 114$~nm, photodissociation proceeds via even more excited electronic states of H$_2$O and systematic quantum calculations are lacking. Time-of-flight spectroscopy indicates that at $\lambda = 115$~nm OH distributions are vibrationally hotter than those computed through the $\tilde{B}$ but still rotationally hot \citep{2019NatCo..10.1250C}. Photodissociation shortward of $\lambda = 114$~nm may also produce preferentially atomic oxygen instead of OH. In this work, \LE{we neglected} the contribution of OH produced through photodissociation at wavelengths shorter than 114~nm.

Our adopted chemical dataset pertains only to water in the rotational ground state ("cold water"). Experimental data suggest that photodissociation of "warm water" could result in a small change in the rotational distribution of OH in the range $N\simeq 32-40$ \citep{1999JChPh.110.4123H}.
This change would however affect only lines coming from $N \gtrsim 32$ in the $9-10.5~\mu$m range (see Sec. \ref{sec:results}). Additionally, \LE{the adopted dataset does} not include the fine-structure and the $\Lambda$-doubling states of the OH products. Recent theoretical investigations suggest that photodissociation through the $\tilde{B}$ state produces OH preferentially in the $A'$ symmetric states, that corresponds to the ($\Omega = 1/2, e$) and ($\Omega = 3/2, f$) states \citep{2015JChPh.142l4317Z}. 
This asymmetry is also expected to depend on the rotational state of the parent H$_2$O, which is not considered here. New quantum calculations resolving the fine-structure and the $\Lambda$-doubling state and including the rotational state of the parent H$_2$O are needed to study in detail the $A'/A''$ asymmetry. Our predictions presented in Sec. \ref{sec:results} remain valid by summing the line intensities over the fine-structure and $\Lambda$ doublets of each $N_{\rm{up}}$ line. 


\subsection{OH excitation}

\BT{When radiative processes dominate, the emitted spectrum of OH will be governed by the state-specific dissociation rates of H$_2$O (see Eq. (\ref{eq:def-ki})) and by the spontaneous transition probabilities in the subsequent cascade of vibronic and rotational transitions. This is called the prompt spectrum of OH. In astrophysically relevant conditions, excited states of OH can be populated in \LE{steady-state} by various other collisional and radiative processes, and these excited molecules also contribute to an observable spectrum. Thus the true population of excited OH differs from the nascent population.}

In this work, the OH level populations and the line intensities were computed with a new code named {\tt GROSBETA} (Black et al. in prep.). This code is based on a \BT{single-zone} model, following the formalism presented in \citet{2007A&A...468..627V} and implemented in the {\tt RADEX} \footnote{\url{https://home.strw.leidenuniv.nl/~moldata/radex.html}} code. Under the assumption of statistical equilibrium, the local population densities, denoted $n_i$ $[\text{cm}^{-3}]$, are given by
\begin{equation}
\frac{d n_i}{d t} = \sum_{j \neq i} P_{ji} n_j - n_i \sum_{j \neq i} P_{ij} +  \sum_k F_i^k - \sum_{k'} n_i \mathcal{D}_i^{k'} = 0,
\label{eq:statistical-eq}
\end{equation}
where $F_i^k$ and $n_i \mathcal{D}_i^{k'}$ are the formation and destruction rates ($[\text{cm}^{-3} \text{s}^{-1}]$) for level $i= (\Lambda, \varv, N, \Omega, \epsilon)$ and are associated with the chemical reactions labeled $k$ and $k'$, respectively. The $P_{ij}$ are the radiative and collisional transition probabilities $i \rightarrow j$ given by
\begin{equation}
  P_{ij}=\begin{cases}
     A_{ij} + B_{ij} \bar{J}_{\nu_{ij}}+ C_{ij}  & (E_i > E_j)\\
     B_{ij} \bar{J}_{\nu_{ij}} + C_{ij} & (E_i < E_j).
  \end{cases}
  \label{eq:Pij}
\end{equation}
Here, $A_{ij}$ and $B_{ij}$ are the Einstein coefficients of spontaneous and induced emission, $C_{ij}$ are \revbis{the collisional rates [s$^{-1}$]}, and $\bar{J}_{\nu_{ij}}$ is the mean specific intensity at the frequency of the radiative transition $i \rightarrow j$ averaged over the line profile. As in \citet{2007A&A...468..627V}, the contribution of the lines to the local radiation field \LE{was computed} following the escape probability method for \rev{a uniform sphere} setting the line width to $\Delta V = 2$ km s$^{-1}$. When all other physical parameters (e.g., density and temperature) are fixed, the solution depends only on the value of the ratio $N($OH$)/\Delta V$.

\BT{The collisional (de)excitation rates in Eq. (\ref{eq:Pij}) \LE{were} computed considering collision with H$_2$ and He for the OH levels within the $\varv = 0$ state using the rate coefficients from \citet{1994JChPh.100..362O} and \citet{2007CPL...445...12K}, respectively.} The rate coefficients computed by \citet{1994JChPh.100..362O} that include levels up to $N=5$ have been extrapolated to higher $N$ numbers (see Appendix \ref{app:collisional-rates-coeff}). 

The induced radiative rates in Eq. (\ref{eq:Pij}) depend upon mean intensities $\bar J_{\nu}$ at the frequencies of all OH transitions. Thus the excitation model requires a specified ambient radiation field. 
Pure rotational and vibration-rotation transitions involve infrared radiation, which \LE{we parametrized} as a blackbody intensity (Planck function) at a temperature \TIR~times a geometrical dilution factor $W$. For illustration, \LE{we took} $W=1$. This gives a simple parametrization of the IR radiation field and a reasonably good proxy for the local radiation field in the far-IR regime relevant for the IR radiative pumping of pure rotational lines of the OH($X^2\Pi$)($\varv=0$) state. \rev{The impact of a more complex and realistic IR radiation field on the line intensities is briefly discussed in Sec. \ref{subsection:low-N-lines}}. Electronic transitions $A-X$ respond to the near-ultraviolet. In this work, \LE{we neglected} the impact of the near-UV radiative pumping.

\rev{The impact of formation and destruction of OH on its level population is described by the last two terms in Eq. (\ref{eq:statistical-eq}). 
When activation barriers can be overcome ($T_{\rm{K}} \gtrsim 200$~K), OH is primarily produced by the reaction of O atoms with H$_2$ and is rapidly converted into H$_2$O by the reaction with H$_2$. In turn, OH can also be regenerated from H$_2$O by UV photodissociation or by the reaction with atomic hydrogen. Photodissociation of OH, which is generally slower than H$_2$O photodissociation \citep[depending on the spectral shape of the UV radiation field, see][]{2017A&A...602A.105H}, can also limit the abundance of OH. \LE{The formation and destruction routes are described by the state specific rates $F_i^k$ and $\mathcal{D}_i^{k'}$, respectively.} Their relative importance depends on the exact physical conditions. A complete model coupling chemistry and excitation of OH is beyond the scope of the present paper and for the sake of generality and conciseness, \LE{we made} a number of simplifying assumptions to focus on the impact of H$_2$O photodissociation on OH excitation in various environments with a limited number of free parameters. The possible impact of O+H$_2$ on the excitation of OH is discussed in Sec. \ref{subsubsec:chemical-pumping}.
In this work, \LE{we assumed} that chemical steady-state holds, so that
\begin{equation}
F \equiv \sum_k \sum_i F_i^k =  \sum_i n_i \mathcal{D}_i,
\end{equation}
where $F$ is the total formation rate of OH ([cm$^{-3}$ s$^{-1}$]), and $\mathcal{D}_i \equiv \sum_k \mathcal{D}_i^k$ the total destruction rate of level $i$ ([s$^{-1}$]). \LE{We also neglected} any state selective destruction of OH assuming that OH is destroyed at the same rate for each state denoted here as $\mathcal{D}$ (i.e., $\mathcal{D}_i$ does not depend on $i$)\footnote{$\mathcal{D}_i$ is the probability of OH in a given state to be destroyed [in s$^{-1}$]. Consequently, this assumption leads to $\mathcal{D} = \mathcal{D}_i$ and not $\mathcal{D}_i = \mathcal{D}/n$, with $n$ the number of OH levels included in the model, as stated in \citet{2009A&A...505..195S}.}. Finally, \LE{we assumed} that only H$_2$O photodissociation leading to OH modifies the level population of OH, with a state specific rate of $F_i = n($H$_2$O$) k(i)$, where $n$(H$_2$O) is the number density of H$_2$O and $k(i)$ the state specific formation rate defined in Eq. (\ref{eq:def-ki}).}


Equation (\ref{eq:statistical-eq}) can then be rewritten as an equation on the column density of OH in the levels $i$ denoted as $N_i$, upon which line intensities depend. \rev{Assuming constant concentrations and excitation conditions along the line of sight,} this yields
\begin{equation}
\sum_{j \neq i} P_{ji} N_j - N_i \sum_{j \neq i} P_{ij} +  \Phi \times \left( f_i - \frac{N_i}{N(\rm{OH})} \right) = 0,
\label{eq:statistical-eq-red}
\end{equation}
with $f_i$ the state distribution of the OH fragments as defined in Eq. (\ref{eq:def-fi}) and
\begin{equation}
\Phi \equiv N(\text{H}_2\text{O})~\bar{k}
\label{eq:def-Phi}
\end{equation}
the column density of H$_2$O photodissociated per unit time (\rev{[cm$^{-2}$ s$^{-1}$]}). The state distribution $f_i$ depends on the spectral shape of the UV radiation field. In this work, we explore UV radiation fields of various shape that are given in Fig. \ref{app:cross-section-H2O}, including narrow-band and broadband UV spectra.

\BT{Equation (\ref{eq:statistical-eq-red}) shows that $\Phi$ is the relevant parameter that ultimately controls the impact of prompt emission on the OH line intensities.}
\BT{The relation between the strength of the UV radiation field and $\Phi$ depends on its spectral shape.
 For a Ly$\alpha$ UV radiation field
\begin{equation}
\Phi = 1.8 \times 10^{4} G_0  \frac{N(\rm{H}_2\rm{O})}{10^{13} \rm{cm}^2}~\rm{cm}^{-2}~\rm{s}^{-1},
\label{eq:def-Phi-AN}
\end{equation}
and for a Draine ISRF,
\begin{equation}
\Phi = 5.9 \times 10^{3} G_0  \frac{N(\rm{H}_2\rm{O})}{10^{13} \rm{cm}^2}~\rm{cm}^{-2}~\rm{s}^{-1},
\label{eq:def-Phi-AN}
\end{equation}
where $G_0$ is the intensity integrated between 91 and 200~nm, in units of the \citet{1978ApJS...36..595D} radiation field ($2.6\times 10^{-6}$ W m$^{-2}$).} In this work we adopt $\Phi=10^7~\rm{cm}^{-2}~\rm{s}^{-1}$ as a fiducial value and explore a broad range of values (see Table \ref{table:param}). 

$\Phi$ can also be written as  
\begin{equation}
\Phi  \equiv \mathcal{D} N(\text{OH}),
\end{equation}
where $\mathcal{D}$ (\rev{[s$^{-1}$]}) is generally referred to as a destruction rate \citep[see e.g.,][]{2007A&A...468..627V,2009A&A...505..195S}. Since chemical \rev{steady-state} is assumed, $\mathcal{D}$ is also connected to the formation rate of OH. If other chemical reactions participate to the formation and destruction of OH (see Eq. (\ref{eq:statistical-eq})) and if these additional reactions form and destroy OH in proportion to their local population densities, then Eq. (\ref{eq:statistical-eq-red}) still holds but $\mathcal{D}$ would not be associated with a destruction rate. $\mathcal{D}$ should rather be associated with a chemical pumping rate equal to the formation rate of the species by the considered process (in cm$^{-3}$ s$^{-1}$) divided by its particle number density. 

\LE{The line intensities were computed} taking into account optical depth effects under the large velocity approximation. When calculating the line intensities, \LE{we assumed} that the IR continuum background interacting with the gas is not along the same line of sight as the observations are taken. That is, the IR field contributes to the radiative pumping but not to the line formation. 
Throughout this work, \revbis{we present line intensities $I$ in erg s$^{-1}$ cm$^{-2}$, which are sometimes called emergent fluxes. That is, our $I$ is related to the specific intensity in a line $I_{\nu}$ in erg s$^{-1}$ cm$^{-2}$ Hz$^{-1}$ sr$^{-1}$ by $I = 4\pi \int I_{\nu} d\nu$. One can recover the observed line flux by the relation:}
\begin{equation}
F = \frac{I}{4\pi} \Delta \Omega,
\label{eq:flux-intensity}
\end{equation}
where $ \Delta \Omega$ is the solid angle \rev{subtended} by the source.

In other words, the populations and the line intensities of OH \LE{were computed} from statistical equilibrium calculations involving prompt emission, IR radiative pumping and collisional excitation. Our model is controlled by six parameters summarized in Table \ref{table:param}: the spectral shape of the UV radiation field that determines the distribution of nascent OH denoted as $f_i$, the column density of H$_2$O photodissociated per unit time denoted as $\Phi$, the column density of OH $N($OH$)$, \BT{the proton density} $n_{\text{H}}$, the kinetic temperature \TK, and the temperature of the IR radiation field $T_{\rm{IR}}$.

\begin{table}
\caption{Parameters of the model and their fiducial values.}              
\label{table:param}      
\centering                                      
\begin{tabular}{c c c c}          
\hline\hline                        
Parameter & Units  & Range & Fiducial \\    
\hline                                   
Shape of the UV field & . & \it{see text} & Ly$\alpha$\\
$\Phi \equiv N(\text{H}_2\text{O}) k$ & cm$^{-2}$~s$^{-1}$ & $10^{4}$-$10^{10}$  & $10^{7}$ \\
$N(\text{OH})$ & cm$^{-2}$ & $10^{12}-10^{16}$  & $10^{14}$ \\
$n_{\text{H}}$ & cm$^{-3}$ & $10^{5}$-$10^{13}$  & $10^{7}$ \\
$T_{\text{K}}$ & K & $-$  & 500 \\
$T_{\text{IR}}$ & K & 50, 120  & 120 \\
\hline   
$\mathcal{D}^{(1)}$ & s$^{-1}$ & $10^{-12}$-$10^{-2}$  & \rev{$10^{-7}$} \\      
\hline                                            \hline                                            
\end{tabular}
\tablefoot{(1) the chemical pumping rate is not considered as a free parameter here since $\mathcal{D} \equiv \Phi / N($OH$)$.} 
\end{table}

\section{Results}

\begin{figure}
\centering
\includegraphics[width=.48\textwidth]{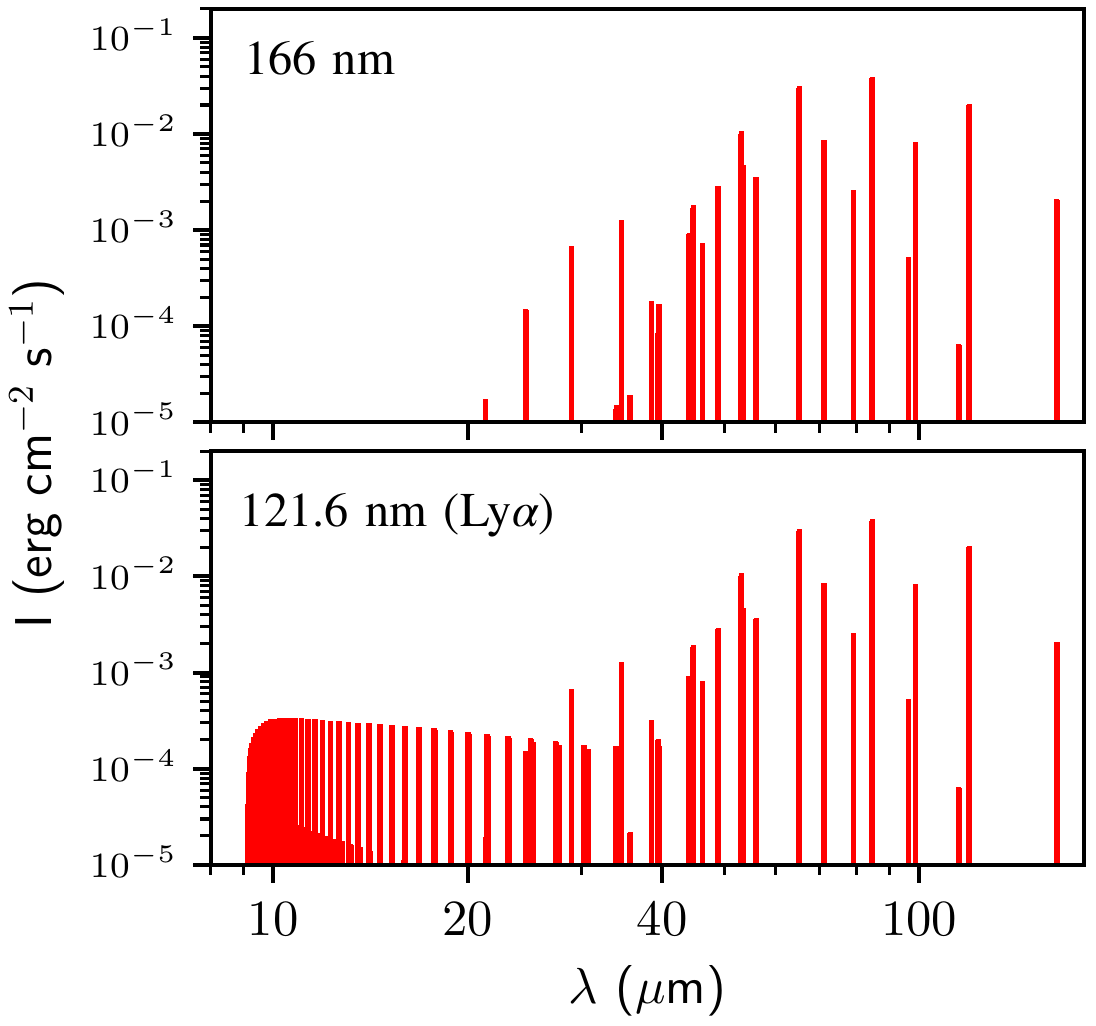} 
\caption{Infrared spectra of OH following H$_2$O photodissociation at two photon wavelengths computed with {\tt GROSBETA}. \textit{Top:} At $\lambda = 166$~nm, \LE{H$_2$O dissociating via its first excited $\tilde{A}$ electronic state and producing rotationally cold OH.} The OH infrared spectrum is dominated by far-IR lines excited by collisions and IR radiative pumping. \textit{Bottom:} \LE{At $\lambda = 121.6$~nm, H$_2$O dissociating via its second excited $\tilde{B}$ electronic state and producing rotationally hot OH.} OH lines in the mid-IR with $N_{\text{up}} = 20-45$ trace H$_2$O photodissociation through this electronic state.
The density, temperature of the background radiation and kinetic temperature are fixed to their fiducial values (see Table \ref{table:param}) and $\Phi = 10^{10}$~cm$^{-2}~$s$^{-1}$. In this figure, all lines are colored red, even if coming from cross-ladder and intra-ladder transitions.}
\label{fig:OH-spectra-intro}
\end{figure}

Figure \ref{fig:OH-spectra-intro} illustrates the difference between OH spectra following H$_2$O photodissociation through the $\tilde{A}$ and $\tilde{B}$ state. Photodissociation via the H$_2$O $\tilde{A}$ state leads to an OH spectrum dominated by far-IR lines coming from low rotational levels. H$_2$O photodissociation does not impact the mid- and far-IR spectrum since only collisions and IR pumping contribute to the excitation of those lines. In contrast, photodissociation through the H$_2$O $\tilde{B}$ state produces additional lines lying in the mid-IR coming from high-$N$ states ($15 \lesssim N \lesssim 45$). These lines are produced by H$_2$O photodissociation. In this section, we shall identify the physical quantities that can be retrieved from the intensities of the mid- and far-IR lines. To do so, an in-depth study of the excitation mechanisms is provided in the case of photodissociation by Ly$\alpha$ photons (121.6~nm). The impact of a broadband UV radiation field, for which photodissociation proceeds through both the H$_2$O $\tilde{A}$ and $\tilde{B}$ states is then studied.

\label{sec:results}

\subsection{Lyman alpha}
\label{subsec:ly-alpha}

\begin{figure*}
\centering
\includegraphics[width=1.\textwidth]{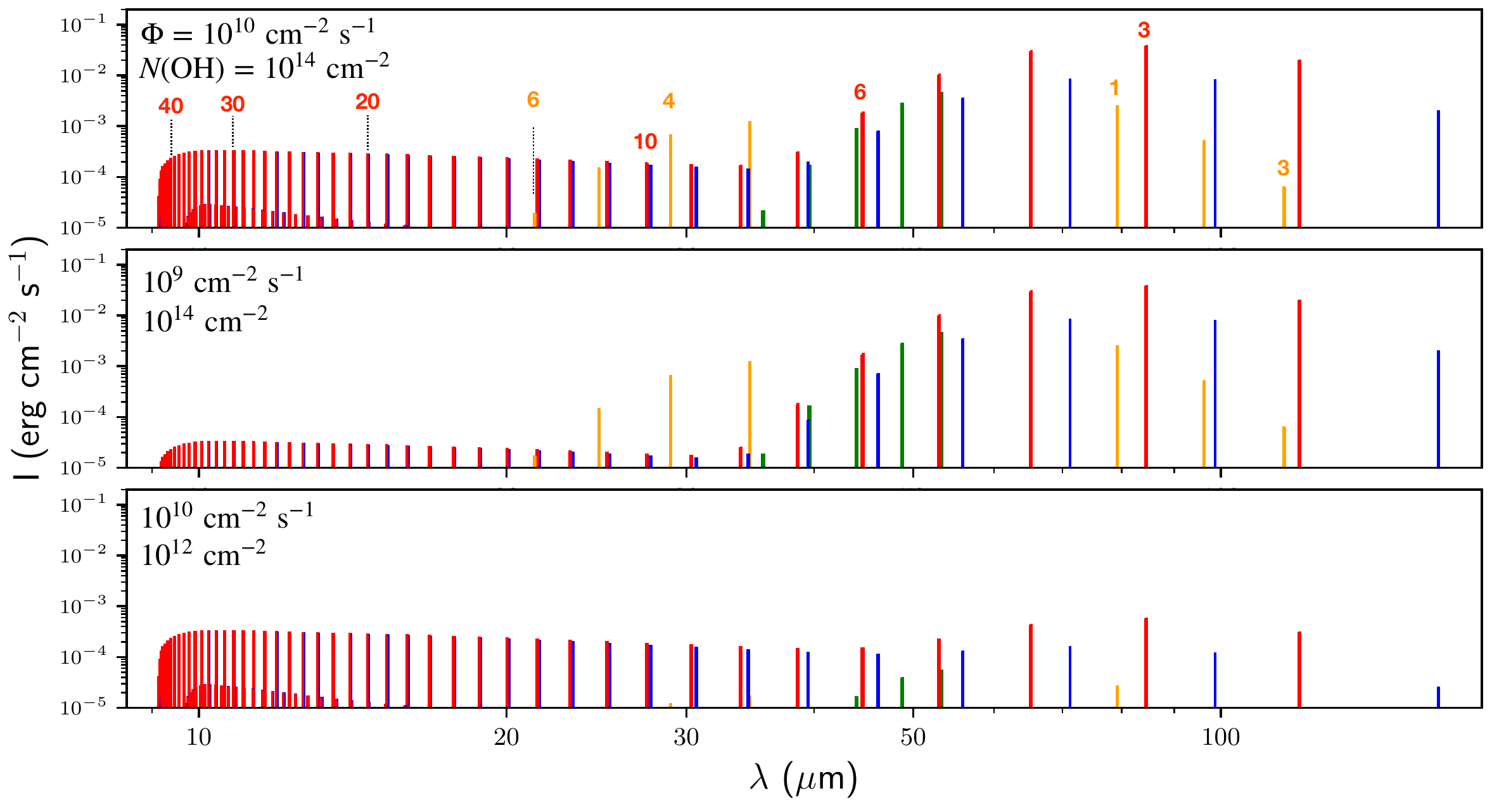} 
\caption{Infrared spectra of OH depending on the column density of H$_2$O photodissociated per unit time $\Phi$ and on $N($OH$)$ computed with {\tt GROSBETA} for a Ly$\alpha$ UV radiation field. The density, temperature of the background radiation and kinetic temperature are fixed to their fiducial values (see Table \ref{table:param}). Following the color code used in Fig. \ref{fig:OH-rot-structure}, intra-ladder rotational lines are in red and blue and cross-ladder lines are in orange and green. $\Lambda$-doubling lines are too weak to appear here. The weak mid-IR lines lying in the range $\sim$ 10-13~$\mu$m are intra-ladder rotational lines within the OH($X$)($\varv=1$) state. The $N_{up}$ rotational number of the upper energy level is indicated for selected lines.}
\label{fig:OH-spectra-ex}
\end{figure*}


\begin{table}
\caption{Rotational lines used as a template for the different excitation regimes.}              
\label{table:line}      
\centering                                      
\begin{tabular}{c c  c c c}          
\hline\hline                        
Transition & $\lambda_{ij}$  & $E_{i}$  & $A_{ij}$ \\    
  $^2\Pi_{\Omega'}(N',\epsilon') \rightarrow $ $^2\Pi_{\Omega''}(N'',\epsilon'')$    & ($\mu$m)  &  (K)   & (s$^{-1}$) \\
\hline                                   
$^2\Pi_{1/2}(30,f) \rightarrow$ $^2\Pi_{1/2}(29,f)$  & 10.8  & 22600  & 3.5(2) \\
$^2\Pi_{1/2}(5,f) \rightarrow$ $^2\Pi_{3/2}(3,f)$ & 24.6 & 875  & 3.8(-2) \\

$^2\Pi_{3/2}(10,f) \rightarrow$ $^2\Pi_{3/2}(9,f)$ & 27.4 & 2905  & 2.0(1) \\
\hline   
\end{tabular}
\end{table}

\begin{figure*}
\centering
\includegraphics[width=0.86\textwidth]{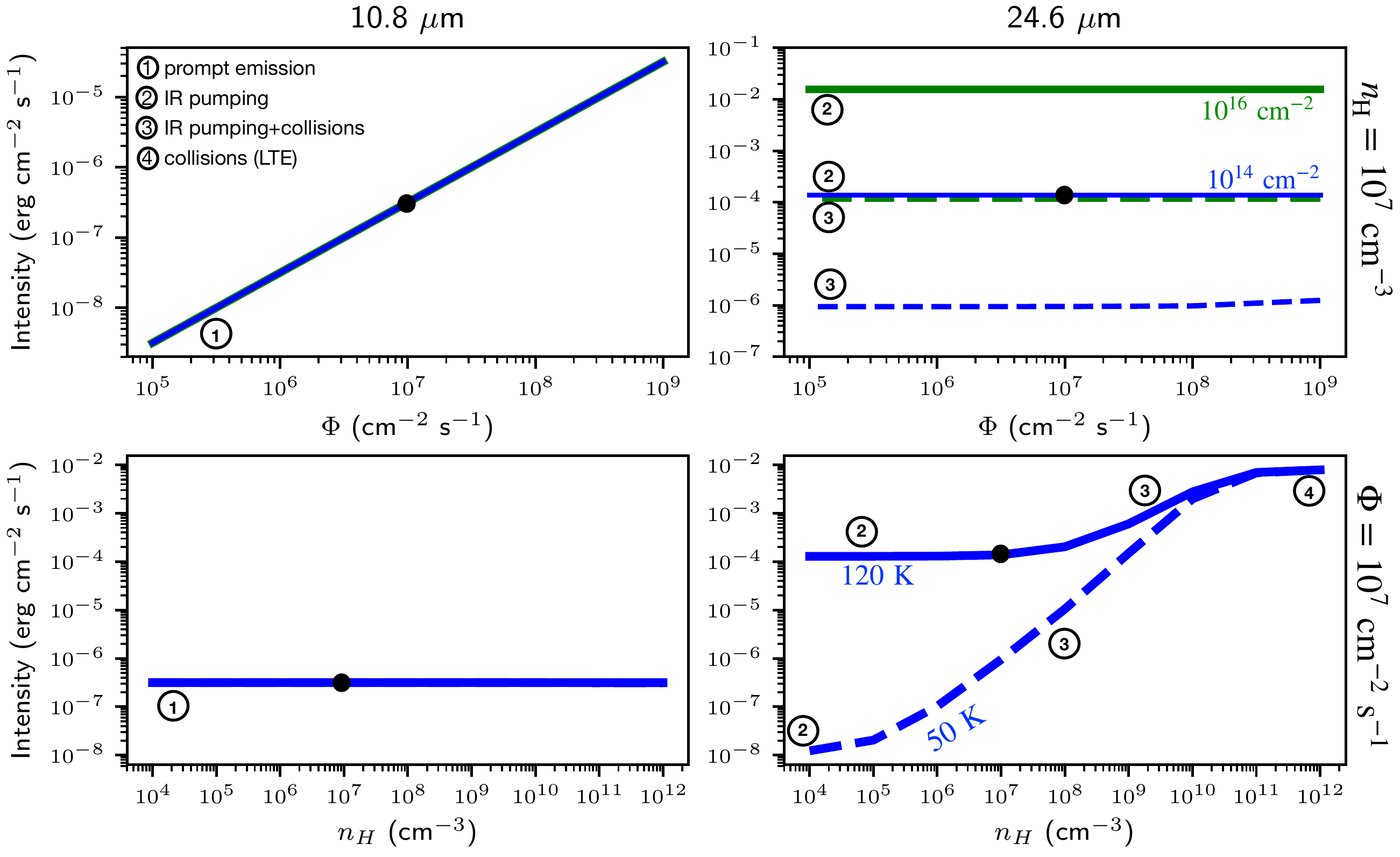} 
\caption{OH line intensities and the associated excitation processes as a function of $\Phi$ and \nH~for various values of $N($OH$)$ and $T_{\rm{IR}}$. The color indicates the value of $N($OH$)$ and the line style the value of \TIR~as defined in the right panels. The black circle indicate the fiducial model. The processes that dominate the excitation of the lines depending on the explored parameters are indicated along each curve. \textit{Left:} Intra-ladder rotational line at 10.8~$\mu$m coming from a high-$N$ level ($N=30$, $E_{\text{up}} = 22600$~K). This line depends only on $\Phi$. \textit{Right:} Cross-ladder rotational line at 24.6~$\mu$m coming from a low-$N$ level ($N=6$, $E_{\text{up}} = 875$~K). This line traces the bulk population of OH. As such, it depends on $N($OH$)$, $T_{\rm{IR}}$ and \nH~but does not depend on $\Phi$.}
\label{fig:lines-lya}
\end{figure*}

As seen in Sect. \ref{subsec:chem-dataset}, photodissociation of H$_2$O by Ly$\alpha$ photons produces OH in high rotational states (Fig. \ref{fig:example-distrib}-b). Figure \ref{fig:OH-spectra-ex} \LE{illustrates the influence of the column density of H$_2$O photodissociated per unit time denoted as $\Phi$} and of the column density of OH, $N($OH$)$. The results of a more systematic exploration of the parameter space are presented in Fig. \ref{fig:lines-lya} by focusing on the intensities of three representative rotational lines that will be observed by JWST-MIRI (see Table \ref{table:line}).


\subsubsection{High-$N$ lines: Prompt emission}
\label{subsec:mid-IR}

\begin{figure}
\centering
\includegraphics[width=.48\textwidth]{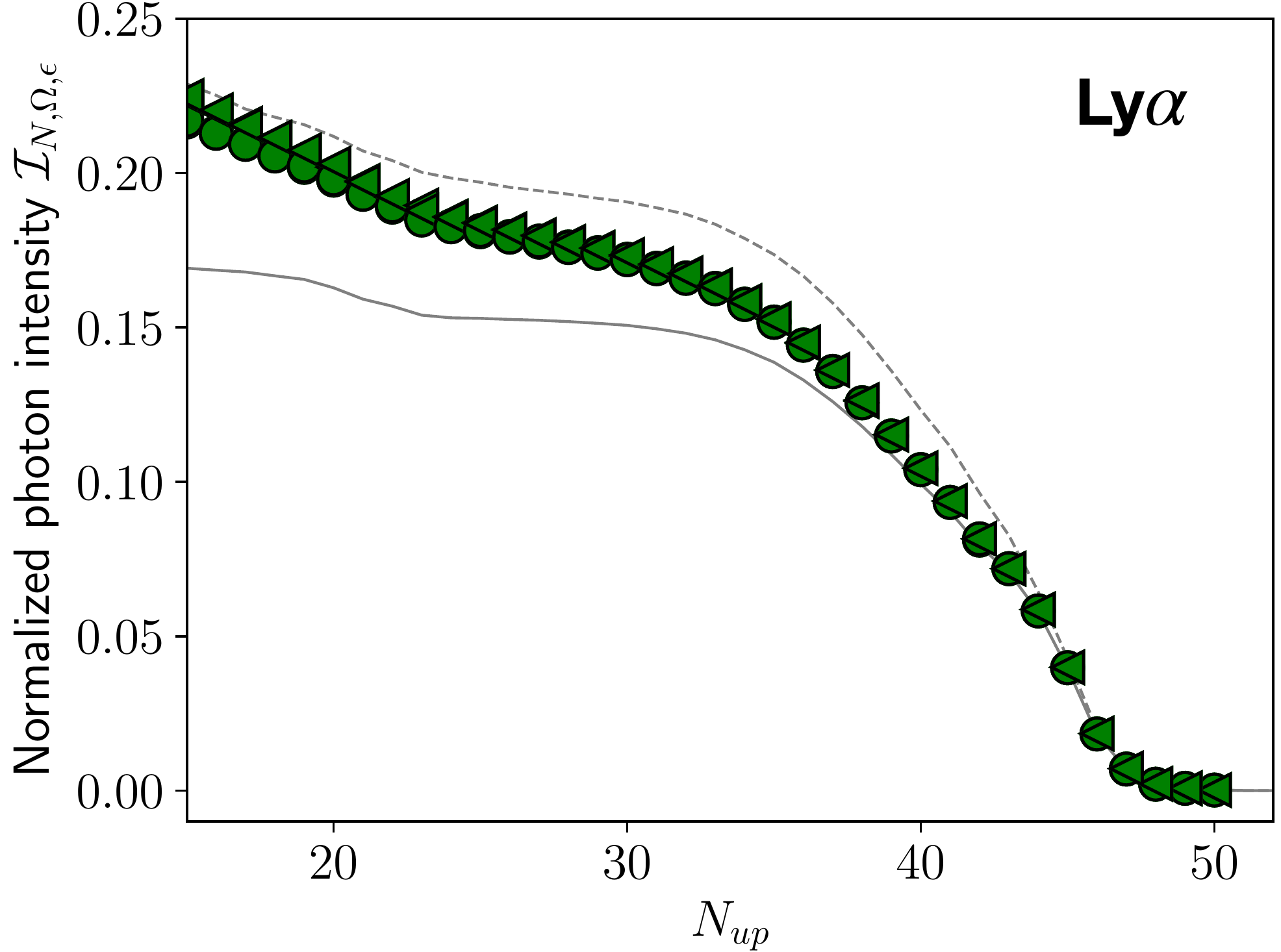}
\caption{Normalized photon intensity of the OH($X^2\Pi$)($\varv=0$) intra-ladder mid-IR lines as a function of the quantum number of the upper energy level $N_{\text{up}}$. The photon intensities are computed for the fiducial values of the parameters and divided by $\Phi$ (see Eq. (\ref{eq:normalised-int})). Circle and triangle markers correspond to lines belonging to the $\Omega = 1/2$ and $\Omega = 3/2$ ladders, respectively. $\Lambda$-doublets are indiscernible in this plot. In this regime, rotational levels are only populated by the radiative cascade of OH photofragments and $\mathcal{I}_{N,\Omega,\epsilon}$, depends only on the spectral shape of the adopted radiation field. 
The solid line is an analytical model that assumes that the intra-ladder lines within the $\varv=0$ state  are powered by the radiative cascade of the OH fragments produced in the $\varv=0$ state. The dotted line is an analytical model that assumes that any OH produced in a vibrational state instantaneously decays toward the ground vibrational state with negligible change in its rotational number (see Appendix \ref{app:anal-model}).} 
\label{fig:lines-lya-mid-IR}
\end{figure}

The mid-IR spectrum ($\lambda \lesssim 20~\mu$m) is dominated by pure intra-ladder rotational lines emerging from levels with high rotational quantum number ($N \ge 14$), corresponding to upper energies $>5000$~K. Figure \ref{fig:OH-spectra-ex} shows that in this wavelength range, the line intensities are higher for a larger value of $\Phi$ (top and middle panels) but do not depend on \NOH~(top and bottom panels). The relative intensities of the intra-ladder lines depend neither on $N($OH$)$ nor on $\Phi$ so we focus on the intensity of the $^2\Pi_{1/2}(30,f) \rightarrow$ $^2\Pi_{1/2}(29,f)$ line at 10.8~$\mu$m (see Table \ref{table:line}) as a proxy for the intensities of the mid-IR lines. Figure \ref{fig:lines-lya} (left panels) shows that the absolute line intensity is directly proportional to $\Phi$ and does not depend on other parameters such as \NOH, \TIR~or \nH. This is one of the most fundamental properties of the high-$N$ lines that makes them an unambiguous diagnostic of H$_2$O photodissociation.

This very simple result points toward a simple excitation process. Due to the high energy of the upper levels, IR radiative pumping does not contribute to the excitation of these lines. De-excitation by stimulated emission by the IR background is also negligible as long as the photon occupation number is much smaller than unity in the mid-IR domain, a condition that is fulfilled over the full parameter space. Due to the very high critical densities of these levels ($n_{\rm{crit}} \gtrsim 10^{13}$~cm$^{-3}$), collisional (de)excitation is also negligible. Instead, the level populations, and the intensity of the lines coming from those levels, are set by the radiative cascade following the formation of OH in high-$N$ states by H$_2$O photodissociation. The line intensities, or equivalently the number of radiative transitions per unit time between two excited levels, are then directly proportional to the formation rate of OH in higher excited states, which is proportional to the photodissociation rate of H$_2$O. The proportionality between the mid-IR line intensity and $\Phi$ is thus a direct consequence of the radiative cascade.

In order to analyze the relative intensity of the mid-IR lines, it is then convenient to define the normalized and dimensionless line intensity 
\begin{equation}
\mathcal{I}_{N,\Omega, \epsilon} = \frac{I^{\Omega, \epsilon}_{N \rightarrow N-1}}{h \nu \Phi},
\label{eq:normalised-int}
\end{equation}
where $I^{\Omega, \epsilon}_{N \rightarrow N-1}$ is the integrated intensity (in erg s$^{-1}$ cm$^{-2}$) of the $N \rightarrow N-1$ line within the ladder $\Omega$ and the parity $\epsilon$, and $\nu$ is the frequency of the line. As long as the population of the high-$N$ levels is set by the radiative cascade, $\mathcal{I}_{N,\Omega, \epsilon}$ depends only on the distribution of nascent OH, or equivalently on the spectral shape of the UV field, and not on $\Phi$ or on any other physical parameter. Figure \ref{fig:lines-lya-mid-IR} shows that $\mathcal{I}_{N,\Omega, \epsilon}$ depends mostly on $N$ and very little on $\Omega$ and $\epsilon$. This is due to the fact that in our model, levels are assumed to be populated by H$_2$O photodissociation regardless of their $\Omega$ and $\epsilon$ states. In the following, we thus omit the reference to $\Omega$ and $\epsilon$ and note the normalized line intensity as $\mathcal{I}_N$. $\mathcal{I}_N$ increases with decreasing $N$, with a stiff rise between $N=47$ and $N=35$. This feature is also visible in Fig. \ref{fig:OH-spectra-ex} where the line intensities increase with wavelength between $9$ and $10~\mu$m. 
For $22 \le N \le 32$, $\mathcal{I}_N$ is rather constant, corresponding also to a rather flat mid-IR spectra between 10 and 20~$\mu$m (see Fig. \ref{fig:OH-spectra-ex}). This results in suprathermal excitation temperatures that vary between \BT{2300~K} for the lines at $\lambda \simeq 20~\mu$m up to \BT{13000~K} for the lines at $\lambda \simeq 10~\mu$m.

The normalized intensity $\mathcal{I}_N$ is directly related to the distribution of nascent OH. Interestingly, $\mathcal{I}_N$ can be interpreted as the probability that a photodissociation event H$_2$O $\rightarrow$ H + OH eventually leads to a radiative decay $N\rightarrow N-1$ via the radiative cascade. As such, $\mathcal{I}_N$ is necessarily smaller than unity. Because we assume the spin-orbit and $\Lambda$-doubling states to be equally populated by H$_2$O photodissociation, $\mathcal{I}_N \lesssim 1/4$. Owing to the selection rules, the rotational cascade within the $\varv = 0$ state is dominated by $N \rightarrow N-1$ transitions. Moreover, OH produced in an electronic level is shown to rapidly decay toward OH($X^2\Pi$) with little change of rotational number. Consequently, the formation of an OH fragment in an OH($\Lambda, \varv=0,N'$) state eventually leads to an intra-ladder transitions $N \rightarrow N-1$, with $N \le N'$. We plot in Fig. \ref{fig:lines-lya-mid-IR} (solid line) an analytical prediction of $\mathcal{I}_N$ assuming that transitions $N \rightarrow N-1$ are supplied by the radiative decay of the OH fragments produced in the $\varv = 0$ states (see model in Appendix \ref{app:anal-model}). This analytical expression relies only on the distribution of nascent OH denoted as $f_i$. The model reproduces the increase in $\mathcal{I}_N$ with deceasing $N$ well, showing that its global variation with $N$ is mostly due the fact that more and more OH fragments are added to the $N \rightarrow N-1$ rotational cascade as $N$ decreases. The radiative cascade should then rather be seen as a \LE{''radiative river''} that grows by it tributaries. In particular, the steep increase in $\mathcal{I}_N$ from $N=47$ to $35$ is due to the fact that most of the OH fragments are produced with these $N$-quantum numbers (see Fig. \ref{fig:example-distrib}-b). 

We also note that for $N<35$, our analytical model progressively underestimates $\mathcal{I}_N$. This is due to the contribution of the OH fragments produced in vibrationally excited states that are not included in our first analytical model. In Fig. \ref{fig:lines-lya-mid-IR} (dashed line) we show the analytical prediction of $\mathcal{I}_N$ assuming that any OH produced in an OH($\Lambda$)($\varv,N$) state immediately decays toward the OH($X^2\Pi$)($\varv=0,N-1$) state. The model overestimates $\mathcal{I}_N$ showing that vibrationally excited states tend to undergo rotational transitions within $\varv \ge 1$ vibrational states before decaying to the $\varv = 0$ state. 

In other words, the spectral shape of the normalized intensity $\mathcal{I}_N$ is set by the population of nascent OH following photodissociation of H$_2$O. Our two simple analytical models, which rely only on the knowledge of the distribution of the OH fragments $f_{\text{i}}$, allow one to bracket $\mathcal{I}_N$.

\subsubsection{Low-$N$ lines}
\label{subsection:low-N-lines}
Figure \ref{fig:OH-spectra-ex} shows that, in contrast to the mid-IR lines, the far-IR lines ($\lambda > 40~\mu$m) \LE{do not depend on the column density of H$_2$O photodissociated per unit time} $\Phi$ but on \NOH. The same conclusions apply to the cross-ladder transitions apparent from 20$~\mu$m to 114$~\mu$m. All these lines arise from $N \lesssim 6$ levels ($E_{\rm{up}} \lesssim 1200$~K) and trace the bulk population of OH that is not excited by the radiative cascade from high-$N$ levels. Because of the high optical depth of the intra-ladder lines, we focus in the following on the optically thin cross-ladder transition at $24.6~\mu$m that arises from a $N=5$ level (see Table \ref{table:line}). Figure \ref{fig:lines-lya} (right) shows for a broader range of parameters that the intensity of this line does not depend on $\Phi$ but on \NOH, \TIR, and \nH. Prompt emission is always negligible and the line intensity is the result of a competition between collisional (de-)excitation and IR radiative pumping.

Figure \ref{fig:lines-lya} (lower right panel) highlights three distinct excitation regimes as a function of density. At low density, the intensity does not depend on the density but on \TIR. In this regime, labeled by \ding{173}, levels are exclusively populated by IR radiative pumping. Because of our specific choice of the IR radiation field, all low-energy levels are thermalized to the same excitation temperature equal to $T_{\text{IR}}$ and the intensities of the optically thin lines are
\begin{equation}
I_{i \rightarrow j} \simeq A_{ij} g_{i} \frac{N(\text{OH})}{Q(T_\text{IR})} \frac{h c}{\lambda_{ij}} e^{-E_i/k_B T_{\text{IR}}},
\label{eq:low-N-intensity}
\end{equation}
where $Q(T_\text{IR})$ is the partition function of OH. The intensity of the line at $24.6~\mu$m is simply proportional to \NOH~and increases with \TIR. For an IR radiation field that deviates from a blackbody, the excitation of the OH($X^2\Pi$)($\varv=0,N$) levels is more complex. \rev{In this general case, the radiation brightness temperature $T_{\rm{rad}}$ varies with wavelength across the far-IR spectrum of OH ($\lambda \lesssim 120~\mu m$). As a rule of thumb, the line intensity is then bracketed between our predictions with an undiluted blackbody at a value of \TIR~ that lies between the minimum and maximum values of $T_{\rm{rad}}$.}

At intermediate density, the intensity increases with \nH. For our fiducial values of \TIR~and \TK, it corresponds to densities between 10$^8$ and 10$^{10}$\cmsq. In this regime, labeled by \ding{174}, collisions contribute to the excitation of the levels. However, line intensities also depend on \TIR, showing that IR radiative pumping is also relevant. Since the density is smaller than the critical density of the upper energy level (\nH$\sim 10^{11}$\cmsq), the de-excitation of the level is via radiative decay. Interestingly, the critical density above which collisions contribute to the excitation of the levels depends on \TIR~and on \TK. It is lower for a lower \TIR (Fig. \ref{fig:lines-lya}, bottom right) or for a higher \TK~(not shown here). By comparing the rates of collisional excitation with the IR radiative pumping rate, collisions take over from IR radiative pumping in the excitation of a given $N$-level for
\begin{equation}
n_{\text{H}} \gtrsim n_{\gamma}(\nu_{\rm{N}}) ~n_{\text{crit}}~e^{h \nu_{\rm{N}}/k_B T_{\rm{K}}} \simeq n_{\text{crit}}~e^{h \nu_{\rm{N}}/k_B (1/T_{\rm{K}}-1/T_{\rm{IR}})} ,
\label{eq:n_crit}
\end{equation}
where $\nu_N$ denotes the frequency of the $N \rightarrow N-1$ radiative transition, \rev{$n_{\gamma}$ is the photon occupation number}, and where we assume \rev{$h \nu_{\rm{N}} \gg T_{\rm{IR}}$} for the second term. With $n_{\text{crit}}$ having a weak dependence on $T_{\rm{K}}$, the critical density above which collisions take over from IR radiative pumping depends on the contrast between the temperature of the gas and that of the radiation field.

Finally, above the critical density of the upper energy level ($n_{\text{crit}} \simeq 10^{11}$\cmsq), the line intensity converges toward its LTE value. In this regime, labeled by \ding{175}, collisions control the excitation and the de-excitation of the level and Eq. (\ref{eq:low-N-intensity}) gives the intensity of the optically thin line by substituting \TIR~by \TK.

One has to keep in mind that when IR radiative pumping is relevant for the excitation of a level, the geometry of the source turns out to be of great importance for the formation of the line coming from this level. We recall that the line intensities presented in this work are computed assuming that the IR background does not contribute to the line formation process. If the background IR field does contribute to the observed emission, lines can be weaker or seen in absorption \citep[see OH lines observed by][]{2010A&A...521L..36W,2013A&A...552A..56W}. Collisions can also play a role in the formation of the line even when negligible in the excitation of the levels. Models incorporating specific source geometries are beyond the scope of this work and have already been explored to analyze low-$N$ OH lines observed by \textit{Herschel} toward protostars and photodissociation regions \citep{2011A&A...530L..16G,2013A&A...552A..56W}.

\subsubsection{Intermediate-$N$ lines}
\label{subsubsec:transition-mid-far}

\begin{figure*}[!h]
\centering
\includegraphics[width=1.\textwidth]{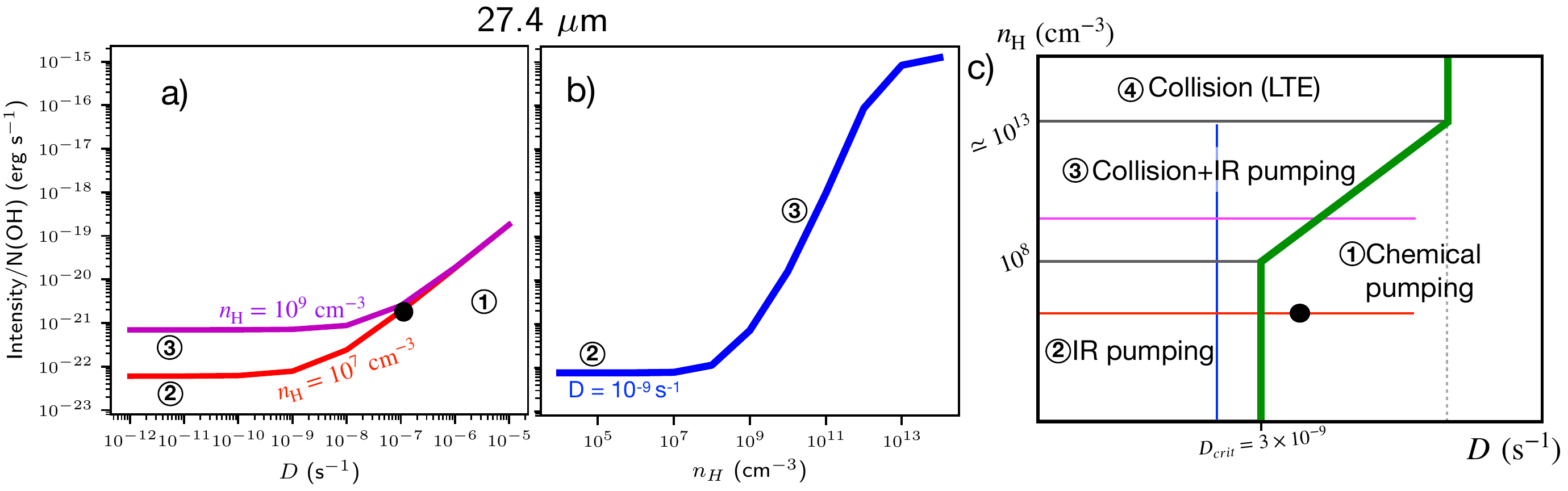}
\caption{Transition from prompt emission to thermal \rev{or radiative excitation} as revealed by the intensity of the intermediate $N=10$ rotational line at 27.4~$\mu$m. The dominant excitation processes are indicated by \ding{172},\ding{173},\ding{174},\ding{175} as defined in Fig. \ref{fig:lines-lya}. a) Line intensity normalized by \NOH~as a function of $\mathcal{D} \equiv \Phi/N(\text{OH})$. The transition between radiative pumping and prompt emission depends on $\mathcal{D}$ and on \nH. In particular, one can define a critical pumping rate $\mathcal{D}_{\rm{crit}}$ below which thermal or radiative excitation processes take over from prompt emission. b) Line intensity normalized by \NOH~as a function of \nH~ for $\mathcal{D} = 10^{-9} \text{s}^{-1} < \mathcal{D}_{\rm{crit}}$. c) Schematic parameter space indicating the dominant excitation process \rev{of the $N=10$ line} as a function of \nH~and $\mathcal{D}$. \rev{We note that whereas the specific values of $\mathcal{D}$ and \nH~reported in the x- and y-axis are for the $N=10$ line, this schematic view remains valid for any rotational line.} The prompt emission region is delimited by $\mathcal{D}_{\rm{crit}}$ (green line). The regime \ding{172} corresponds to the excitation regime  of high-$N$ line whereas the regimes \ding{173}, \ding{174} and \ding{175} correspond to the excitation regime of low-$N$ lines. Panels a) and b) are cuts in the parameter space and are indicated by red, magenta and blue lines. \BT{The temperature of the IR background, the kinetic temperature and the column density of OH are fixed to their fiducial values. The other parameters are indicated in each panel.}}
\label{fig:OH-spectra-transition}
\end{figure*}

We have shown in Section \ref{subsec:mid-IR} that the high-$N$ rotational lines excited by prompt emission are proportional to $\Phi$ and can thus be used to probe the photodissociation of H$_2$O. In contrast, low-$N$ lines do not depend on $\Phi$ and trace the bulk population of OH. It is thus of a great importance to determine if a line coming from an intermediate-$N$ level is indeed excited by prompt emission or by other processes such as IR radiative pumping or collisions.


Figures \ref{fig:OH-spectra-transition}-a and b show that the intermediate$-N$ line at 27.4$~\mu$m, which originates from a $N=10$ level, shares features of low-$N$ and high-$N$ lines. This complex excitation pattern is the result of the competition between prompt emission, IR radiative pumping and collisions. Figure \ref{fig:OH-spectra-transition}-c summarizes the dominant excitation processes as a function of the density~\nH~and the chemical pumping rate $\mathcal{D} = \Phi/N(\rm{OH})$ for the fiducial values of \TIR~and $N($OH$)$.


For high values of the chemical pumping rate, the line intensity normalized by \NOH~is proportional to $\mathcal{D}$ (Fig. \ref{fig:OH-spectra-transition}-a). This corresponds to the right part of the parameter space shown in Fig. \ref{fig:OH-spectra-transition}-c. In this regime, the excitation of the upper energy level is dominated by prompt emission (regime \ding{172}) and we recover the result obtained for the high-$N$ lines that the line intensity is proportional to $\Phi=\mathcal{D} N($OH$)$ (Sec. \ref{subsec:mid-IR}). Figure \ref{fig:OH-spectra-transition}-a shows that below a critical value of the chemical pumping rate, denoted here as $\mathcal{D}_{crit}$, the intensity does not depend on $\mathcal{D}$ as IR radiative pumping or collisions take over from prompt emission. This corresponds to the left region of the parameter space (Fig. \ref{fig:OH-spectra-transition}-c). The excitation of the line as a function of the density then follows a pattern similar to that found for low$-N$ lines (Fig. \ref{fig:OH-spectra-transition}-b). At low density, the excitation is dominated by IR radiative pumping (regime \ding{173}) whereas at higher density collisions progressively take over (regime \ding{174}). We find that for the fiducial values of \TIR~and \TK, collisions contribute to the excitation of the line above \nH$\simeq 10^{8}$~\cmsq. However, we recall that the effect of collisions on levels with $N\gtrsim 5$ is highly uncertain due to the lack of quantum calculations of collisional rate coefficients for these levels. In other words, the parameter space is divided into two regions: a region dominated by prompt emission for which results found in Sec. \ref{subsec:mid-IR} apply, and a region dominated by other excitation processes for which results found in Sec. \ref{subsection:low-N-lines} apply.

The boundary between the two regions is defined by $\mathcal{D}_{crit}$ (green line Fig. \ref{fig:OH-spectra-transition}-c). Its value depends on the parameters that control the thermal \rev{and radiative} excitation of OH, namely \nH, \TK, and \TIR. For example, Fig. \ref{fig:OH-spectra-transition}-a shows that $\mathcal{D}_{crit}$ increases from $\sim 2 \times 10^{-9}$ to $\sim 2 \times 10^{-8}$~s$^{-1}$ by increasing the density from \nH=$10^{7}$ to $10^{9}$~\cmsq. Since $\mathcal{D}_{crit}$ quantifies the competition between \rev{thermal or radiative} excitation and prompt emission, it also depends on the upper energy level of the line. We derive in Appendix \ref{app:crictical-pumping} simple estimates of $\mathcal{D}_{\rm{crit}}$ as a function of the excitation conditions (\nH, \TIR, and \TK) for any rotational level. In particular, we show that the schematic view of the parameter space proposed in Fig. \ref{fig:OH-spectra-transition}-c remains valid for the low-$N$ lines for which the boundary is shifted to the right, and for high-$N$ lines, for which the boundary is shifted to the left by orders of magnitudes.

\subsubsection{Summary}

The integrated intensities of the lines coming from high-$N$ levels ($N \gtrsim 20$) are proportional to the column density of H$_2$O photodissociated per unit time and do not depend on other physical parameters. The shape of the mid-IR spectrum is only set by the distribution of the nascent OH and we define a normalized intensity of the intra-ladder lines, \LE{denoted as $\mathcal{I}_N$} (Fig. \ref{fig:lines-lya-mid-IR}). The cross-ladder and intra-ladder rotational lines coming from low-$N$ levels ($N \lesssim 6$) are populated by IR radiative pumping or collisions and are thus tracing the bulk population of OH. Intermediate-$N$ lines can be either excited by prompt emission, IR radiative pumping and/or collisions (see Fig. \ref{fig:OH-spectra-transition}-c).

\subsection{Other UV radiation fields}
\label{subsec:impact-RF}

\subsubsection{State distribution of the OH fragments}

\begin{figure}
\centering
\includegraphics[width=0.48\textwidth]{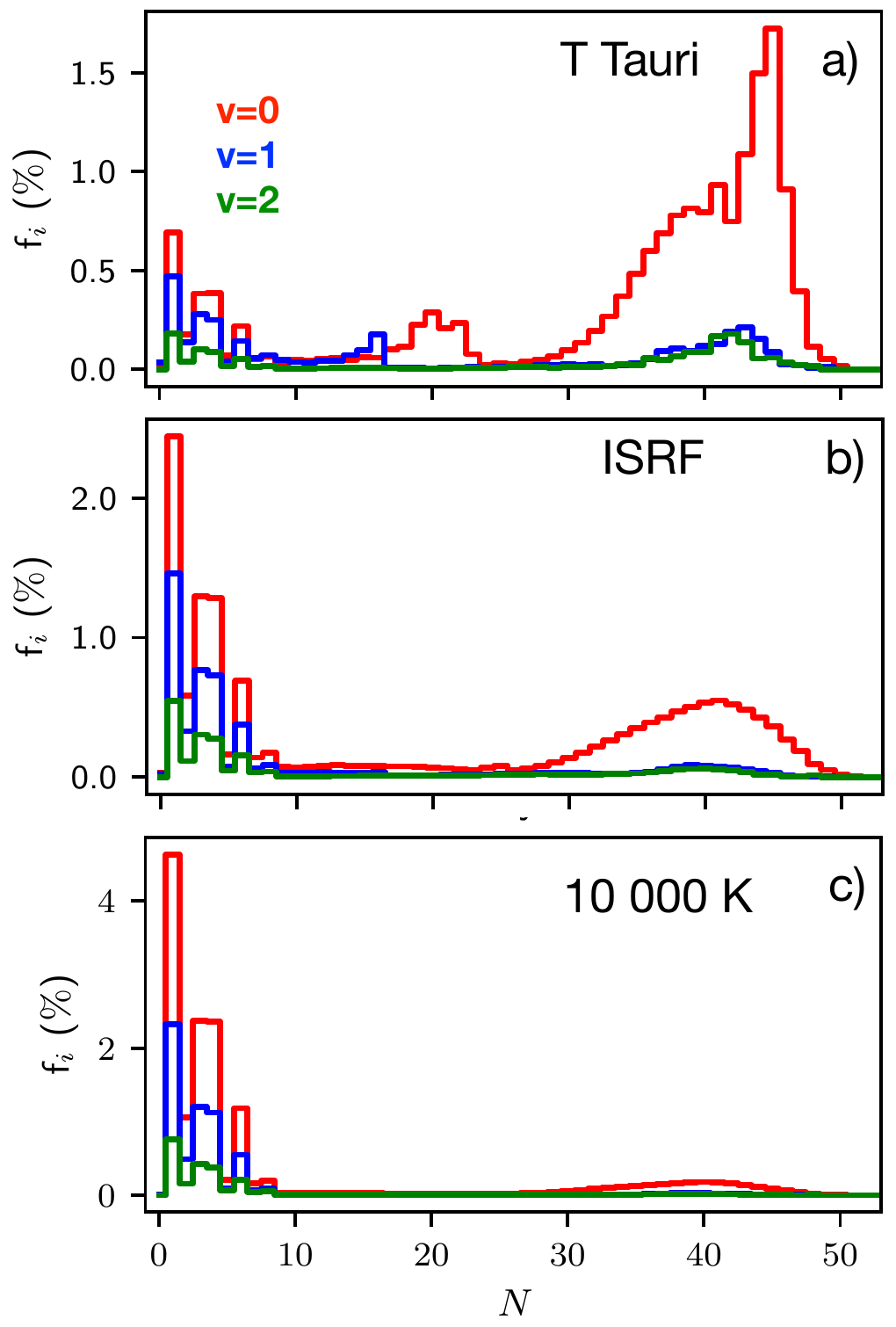} 
\caption{Rotational distribution of the OH fragments following H$_2$O photodissociation by UV radiation fields of different shapes. The impact of three radiation fields is explored: a) a radiation field representative of an accreting T Tauri star dominated by a Ly$\alpha$ emission line, b) an ISRF radiation field and c) a blackbody radiation field at 10 000~K. Vibrational quantum number are color coded as indicated in the top panel. The distribution is summed over the two electronic states. Only vibrational levels that contribute to at least 1.5$\%$ to the population of nascent OH are shown.}
\label{fig:distrib-RF}
\end{figure}

\begin{table}
\caption{Branching ratio between H$_2$O photodissociation \BT{leading to OH} through the $\tilde{A}$ state ($\lambda \gtrsim 143$~nm) and the $\tilde{B}$ state ($ 114 \lesssim \lambda \lesssim 143$~nm) for UV radiation fields of various spectral shapes.}              
\label{table:branching-ratio}      
\centering                                      
\begin{tabular}{c c c}          
\hline\hline                        
Radiation field  &  $\tilde{A}$ & $\tilde{B}$ \\    
\hline                                   
Ly $\alpha$     &  0$\%$& 100$\%$ \\
T Tauri         & 13$\%$&  87$\%$  \\
ISRF            & 46$\%$&  54$\%$  \\
Blackbody 1000K & 82$\%$&  18$\%$  \\
\hline   
\end{tabular}
\label{table:ratio-photo-channel}
\end{table}

Figure \ref{fig:distrib-RF} shows the state distribution of the OH fragments following H$_2$O photodissociation by UV radiation fields of various spectral shapes. As shown in Section \ref{subsec:mid-IR}, the OH($A$)($\varv,N$) states decay toward the ground electronic state with little change of $N$. We consequently plot only the sum of the rotational distribution of OH($X$) and OH($A$) as defined by
\begin{equation}
\tilde{f}_i = f(X,\varv,N)+\frac{1}{2} f(A,\varv,N),
\end{equation}
where the factor $\frac{1}{2}$ stands for the different degeneracies between OH($A$)($\varv, N$) and OH($X$)($\varv, N$) states. The distribution of nascent OH computed from Eqs. (\ref{eq:def-ki}) and (\ref{eq:def-fi}) results from photodissociation at various wavelengths. As shown in Sec. \ref{subsec:chem-dataset}, the distribution of the OH fragment $\eta(\lambda,i)$ depends markedly on the UV wavelength (see Fig. \ref{app:eta}). One of the most prominent differences is between photodissociation \LE{longward of} $\lambda = 143$~nm, that produces OH($X$) in low-$N$ states, and photodissociation \LE{shortward of} this value that produces OH($X$) in high-$N$ states and a small fraction of electronically excited OH($A$) with $ N \le 27$. The distribution of OH following H$_2$O photodissociation by a broad UV radiation field reflects the relative contribution of the different photodissociation channels. The fraction of photodissociation that proceeds through the two channels is given in Table \ref{table:ratio-photo-channel}. 

We first study the effect a radiation field representative of the UV spectrum emitted by an accreting T Tauri star that includes a UV continuum plus emission lines (see Fig. \ref{app:cross-section-H2O}). As shown by \citet{2003ApJ...591L.159B} and \citet{2012ApJ...756L..23S}, Ly$\alpha$ emission dominates over the continuum emission with $\sim 90\%$ of H$_2$O photodissociation done by Ly$\alpha$ photons. It results in a state distribution that is similar to that produced by a pure Ly$\alpha$ radiation field (Fig. \ref{fig:distrib-RF}-a). Contribution of H$_2$O photodissociation through the  H$_2$O $\tilde{A}$ state ($\lambda \ge 143$~nm), that represents $\sim 10 \%$ of the total photodissociation rate, is however seen at $N \le 6$. 

In contrast, a standard UV interstellar radiation field (ISRF) has a smooth and rather flat spectral distribution (Fig. \ref{app:cross-section-H2O}). H$_2$O photodissociation then proceeds through a broad wavelength range and results in a distribution of OH fragments that exhibits features of both photodissociation through the H$_2$O $\tilde{A}$ and $\tilde{B}$ state (Fig. \ref{fig:distrib-RF}-b).
Photodissociation through the $\tilde{B}$ state creates a peak in the rotational distribution around $N = 41$ similar to that produced by photodissociation by Ly$\alpha$, though somewhat smoother. \BT{OH($A$) states are produced with a broad range of rotational quantum numbers, which results in a small bump in the rotational distribution that is perceptible in the range $N=10-27$. We note that a significant fraction of the OH($A$) products are dissociative and therefore not visible in Fig. \ref{fig:distrib-RF}-b.} Photodissociation through the H$_2$O  $ \tilde{A}$ state, that represents $46\%$ of the total photodissociation rate, yields to a prominent peak at low-$N$ numbers. 

A UV radiation field with a blackbody shape at $T=10000$~K has the steepest UV slope. Most of H$_2$O photodissociation proceeds through the $\tilde{A}$ state ($82\%$). The resulting distribution of nascent OH is then dominated by low-$N$ numbers whereas the bump at $N\simeq 40$ is reduced accordingly (Fig. \ref{fig:distrib-RF}-c).

In other words, the state distribution of OH following H$_2$O photodissociation by a broad UV spectrum exhibits a bump at high-$N$ number and a peak at low-$N$ number. The amount of OH produced with high-$N$ numbers is globally proportional to the fraction of photodissociation occurring through the $\tilde{B}$ state ($\lambda < 143$~nm).

\subsubsection{Mid-IR lines}

\begin{figure}
\centering
\includegraphics[width=.48\textwidth]{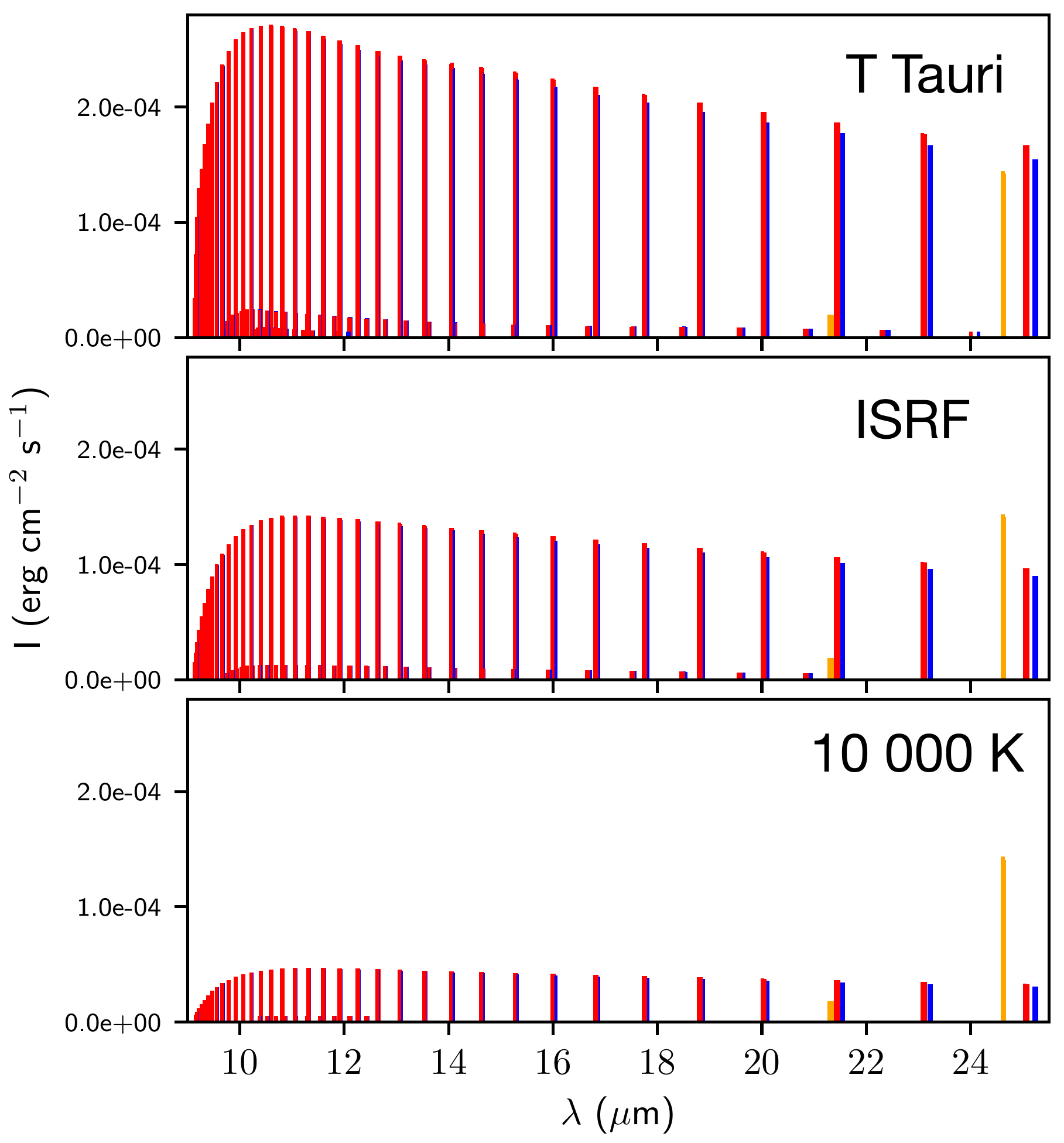}
\caption{OH mid-infrared spectrum for various UV spectra computed with {\tt GROSBETA} for $\Phi = 10^{10}$ cm$^{-2}$~s$^{-1}$. Other parameters are constant and equal to their fiducial values given in Table \ref{table:param}. In this regime, the intensity of intra-ladder lines (red and blue) are proportional to $\Phi$ and do not depend on other parameters such as \TIR~or \nH.}
\label{fig:lines-grid-mid-IR}
\end{figure}

\begin{figure}
\centering
\includegraphics[width=.48\textwidth]{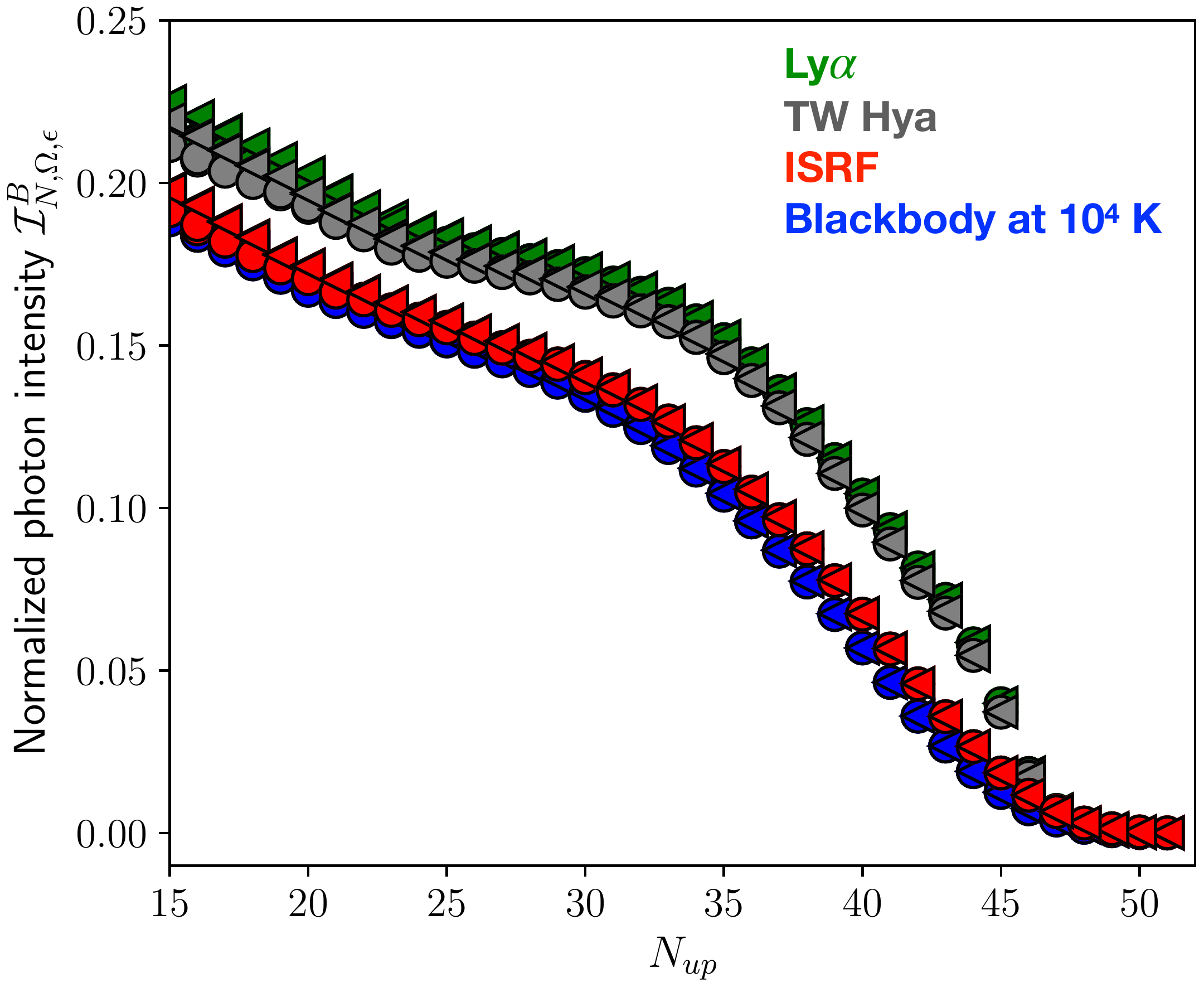}
\caption{Normalized photon intensity of the OH($X^2\Pi$)($\varv=0$) intra-ladder mid-IR lines as a function of the quantum number of the upper energy level $N_{\text{up}}$. The intensities are computed for the fiducial values of the parameters (see Table \ref{table:param}) and normalized by the column density of water photodissociated via its $\tilde{B}$ state (see Eq. (\ref{eq:normalised-int-B})). Circle and triangle markers correspond to lines belonging to the $\Omega = 1/2$ and $\Omega = 3/2$ ladders, respectively. $\Lambda$-doublets are indiscernible in this plot. In this regime, rotational levels are only populated by the radiative cascade of OH photofragments and $\mathcal{I}^{\tilde{B}}_{N,\Omega,\epsilon}$, depends only on the shape of the adopted radiation field. }
\label{fig:measuring-FUV-145nm}
\end{figure}


As shown in the case of H$_2$O photodissociation by a Ly$\alpha$ radiation field, mid-IR lines are proportional to $\Phi$. The same conclusion applies for any UV radiation field and we show in Fig. \ref{fig:lines-grid-mid-IR} the mid-IR spectrum emitted by OH for the UV radiation fields explored in this work, for the same value of $\Phi$.

The mid-IR line intensities depend markedly on the shape of the radiation field. The main difference resides in the absolute line intensity between $10$ and $\sim 20~\mu$m. For example, line intensities are $\sim 5$ times weaker for a blackbody at 10$^4$~K than for a T Tauri radiation field. As seen in the case of a Ly$\alpha$ radiation field, mid-IR lines are fueled by the radiative decay of OH produced in high-$N$ states. The absolute mid-IR line intensities are then proportional to the fraction of OH produced in a rotationally excited state, which is also proportional to the column density of H$_2$O photodissociated through $\tilde{B}$ state \rev{([cm$^{-2}$ s$^{-1}$])} as defined by
\begin{equation}
\Phi^{\tilde{B}} \equiv N(\text{H}_2\text{O}) \int_{114~nm}^{143~nm} \sigma(\lambda) I(\lambda) d \lambda.
\end{equation}
This is further shown in Fig. \ref{fig:measuring-FUV-145nm}, where the line intensities normalized by $\Phi^{\tilde{B}}$ as defined by\footnote{In the case of the photodissociation by a Ly$\alpha$ radiation field studied in Sec. \ref{subsec:ly-alpha}, $\Phi = \Phi^{\tilde{B}}$ and $\mathcal{I}_N= \mathcal{I}^{\tilde{B}}_N$}
\begin{equation}
\mathcal{I}^{\tilde{B}}_{N,\Omega, \epsilon} \equiv \frac{I^{\Omega, \epsilon}_{N \rightarrow N-1}}{h \nu \Phi^{\tilde{B}}}
\label{eq:normalised-int-B}
\end{equation}
are the about same for the different UV fields (within 20$\%$). In other words, the absolute intensity of mid-IR lines traces the amount of H$_2$O photodissociated through the $\tilde{B}$ state ($\lambda < 143$~nm). It follows that the absolute intensity is, to a first approximation, only proportional to $\Phi^{\tilde{B}}$, regardless of the exact shape of the UV radiation field.

\BT{The other difference resides in the relative intensity of the lines, that reveals the precise shape of the rotational distribution of nascent OH fragments. This is best seen in Fig. \ref{fig:measuring-FUV-145nm}. The increase in the line intensities from  $N_{up}=46$ down to $N_{up} \sim 35$ is steeper for T Tauri and Ly$\alpha$ radiation fields than for the other two. These differences are due differences in the exact shape of the rotational distributions of the OH($X$) fragments shown in Fig. \ref{fig:distrib-RF}. In the case of T Tauri and Ly$\alpha$ radiation fields, the peak at high-$N$ is more pronounced than for the broadband UV spectra. The differences in the rotational distributions at lower $N$, that are mostly due to differences in the OH($A$) distributions, result in minor differences in the line intensities in the range $N_{up} = 25-15$.} 

\subsubsection{Low and intermediate-$N$ lines}

Regarding far-IR lines and cross-ladder mid-IR lines, that emerge from $N \lesssim 6$ levels, we find that the integrated intensities depend neither on the shape of the UV radiation field nor on $\Phi$. This is a surprising result since UV radiation fields with a significant flux \LE{longward of} $143$~nm, produce OH with low rotational quantum numbers. Our finding indicates that the contribution of this rotationally cold population of nascent OH to the excitation of low-$N$ levels within $\varv=0$ is negligible compared to infrared pumping or collisions. The only population of nascent OH that affects the intensity of the rotational lines are the one produced with a high rotational number, typically $N \gtrsim 15$. \BT{However, we note that for much lower IR radiation fields and \LE{lower} densities, these lines might be fueled by prompt emission (see Fig. \ref{fig:OH-spectra-transition}-c, regime \ding{172}). In that case, we expect to have a competition pattern between excitation via the H$_2$O $\tilde{A}$ state and the $\tilde{B}$ state}.

As studied in the case of photodissociation by a Ly$\alpha$ radiation field, intermediate-$N$ lines can be excited either by prompt emission or IR radiative pumping and/or collisions. For a given set of physical parameters $\{$\TIR,~\TK, \nH$\}$, the transition between prompt emission and the other excitation processes is controlled by the chemical pumping rate $\mathcal{D} \equiv \Phi/N(\rm{OH})$ as summarized in Fig. \ref{fig:OH-spectra-transition}-c. The analysis proposed in Sec. \ref{subsubsec:transition-mid-far} can be generalized to any UV radiation field by simply substituting $\mathcal{D}$ by $\mathcal{D}^{\tilde{B}}$ as defined by
\begin{equation}
\mathcal{D}^{\tilde{B}} = \Phi^{\tilde{B}} / N(\text{OH}).
\end{equation}
We note that for the explored UV radiation fields, this adaptation has little impact on the transition between prompt emission \rev{and the other excitation processes.}

\section{Discussion and application to HH 211}

\label{sec:disscu}

In this section, we present a method for observationally inferring the photodissociation rate of H$_2$O and deducing the local UV radiation field. As an illustration, our model is applied to the \textit{Spitzer}-IRS observations of the apex of the HH 211 bow-shock published by \citet{2008ApJ...680L.117T}.

\subsection{Diagnostics}

\subsubsection{Column density of H$_2$O photodissociated per second}

Our results show that the absolute intensities of the intra-ladder lines in the mid-IR are proportional to the column density of H$_2$O photodissociated per second via the $\tilde{B}$ state, denoted as $\Phi^{\tilde{B}}$ (see Fig. \ref{fig:measuring-FUV-145nm}). \BT{To our knowledge, the only alternative process that can also excite the rotational levels of energy $E_{\text{up}} \ge 20 000$~K is the electron impact dissociation of H$_2$O \citep{1974CP......6..445B,2019ApJ...885..167B}.
In some situations, such as disks around young stellar objects where there might be a source of energetic electrons ($\gtrsim 10$~eV), the high-$N$ OH lines would thus trace the destruction of H$_2$O via both e-impact and UV photodissociation. Complementary diagnostics, such as lines emerging from the triplet electronic states of H$_2$ or CO that can be excited efficiently by electron impact but not by UV absorption, can help to determine if OH lines might also trace the electron impact dissociation of H$_2$O. In this context, the predictions presented in this work remain valid as long as the destruction rate of H$_2$O by UV photodissociation dominates over destruction by e-impact, a condition that is generally satisfied in irradiated regions.}

Consequently, the mid-IR lines of OH($X$)($\varv=0$) provide a robust measurement of the column density of H$_2$O photodissociated in the range $114$ to $143~$nm. This key quantity can be observationally derived using the relation
\begin{equation}
\Phi^{\tilde{B}} = \frac{I_{N,\Omega, \epsilon}}{h \nu \mathcal{I}^{\tilde{B}}_{N,\Omega, \epsilon}},
\label{eq:measurment-phi}
\end{equation}
with $\mathcal{I}^{\tilde{B}}_{N,\Omega, \epsilon}$ provided in Fig. \ref{fig:measuring-FUV-145nm} and in Appendix \ref{app:conversion}. As shown above, $\mathcal{I}^{\tilde{B}}_{N,\Omega, \epsilon}$ depends little on the spectral shape of the UV radiation field, and, in the absence of any other information, $\Phi^{\tilde{B}}$ can still be derived with a high accuracy ($\sim 10\%$). Interestingly, the time scale associated with the radiative cascade is about 10~ms, which is short compared with relevant dynamical time scales. Therefore, the value of $\Phi^{\tilde{B}}$ should be considered as an instantaneous measurement of the photodissociation activity. If the emitting region is unresolved, the derived value of $\Phi^{\tilde{B}}$ depends on the assumed source size as for any determination of a column density. Strictly speaking, the total flux integrated over a region on the sky gives the total number of H$_2$O molecules photodissociated per second in that region.

\subsubsection{Spectral shape of the UV radiation field}

The small deviations in the relative line intensities could be used to constrain the spectral shape of the radiation field (see Fig. \ref{fig:lines-grid-mid-IR} and \ref{fig:measuring-FUV-145nm}). In particular, our results show that a Ly$\alpha$ dominated radiation field produces a steeper increase in the photon line intensities with decreasing $N_{up}$. However, detecting these small differences in the intensities in the range $9$ to $11~\mu$m requires an accurate relative flux calibration. In fact, the $10~\mu$m silicate feature plays a significant role in the dereddening of the fluxes at visual extinctions $A_{\rm{V}} \gtrsim 10$. In that perspective, the relative intensity of the lines between $\simeq 9$ and $10~\mu$m in regions with high extinction should be interpreted with caution since the exact shape of the extinction curve varies significantly with $A_{\rm{v}}$ \citep{2009ApJ...690..496C,2009ApJ...693L..81M,2020ApJ...895...38H}. We posit that a significant extinction in the mid-IR hampers the diagnostic capabilities of the OH lines to securely constrain the shape of the local UV radiation field. The impact of the rotational excitation of the parent H$_2$O may also complicate the interpretation of the relative line intensities and detailed quantum calculations are needed to propose robust diagnostics based on the exact shape of the OH mid-IR spectrum.

\subsubsection{Photodissociation rate of H$_2$O}

Our work shows that the detection of mid-IR lines of OH is prime evidence for the presence of H$_2$O, in particular in environments where the H$_2$O column density is limited by strong UV fields.
If an independent measurement of the local UV radiation field can be observationally obtained, the column density of H$_2$O that it exposed to the UV field can be deduced as
\begin{equation}
N(\text{H}_2\text{O}) = \frac{\Phi^{\tilde{B}}}{k_{\tilde{B}}},
\end{equation}
where $k_{\tilde{B}}$ is the photodissociation rate in s$^{-1}$ associated with H$_2$O $\rightarrow$ OH + H through the H$_2$O $\tilde{B}$ state.

Conversely, the OH emission can be used to measure the photodissociation rate $k_{\tilde{B}}$ by measuring the column density of water $N(\rm{H}_2\rm{O})$:
\begin{equation}
k_{\tilde{B}} = \frac{\Phi^{\tilde{B}}}{N(\text{H}_2\text{O})},
\label{eq:measurement-kb}
\end{equation}
where $\Phi^{\tilde{B}}$ is measured from the mid-IR lines of OH (see Eq. (\ref{eq:measurment-phi})).
This quantity being a ratio between a column density and $\Phi^{\tilde{B}}$, it does not depend on the assumed source size.

The measurement of $N(\rm{H}_2\rm{O})$ is challenging and may be the major source of uncertainty in the determination of $k_{\tilde{B}}$. Besides the uncertainties that plague the determination of the column densities, the variation of the UV field and of the number density of H$_2$O along the line of sight may be an important bias. Indeed, irradiated environments exhibit a layered structure with physical conditions that vary steeply with distance, due to the attenuation of the radiation field or shocks. Because the OH mid-IR lines are optically thin and depend little on the shape of the UV radiation field, our predictions can be directly extended to any physical structure with
\begin{equation}
I_{N} = \mathcal{I}^{\tilde{B}}_{N} \int_{z} n_{\text{H}_2\text{O}}(z) k_{\tilde{B}}(z) dz,
\label{eq:1D}
\end{equation}
with $n_{\text{H}_2\text{O}}(z)$ the local number density of H$_2$O and $k_{\tilde{B}}(z)$ the local photodissociation rate of H$_2$O through the $\tilde{B}$ state. $\Phi^{\tilde{B}}$ being a quantity integrated along the line of sight, a spatial variation of the UV radiation field and of $n_{\text{H}_2\text{O}}(z)$ does not affect its measurement. On the contrary, $k_{\tilde{B}}$ is a local quantity and Eq. (\ref{eq:measurement-kb}) should be applied with caution. Considerations on the geometry and on the viewing angle of the source may also help.


\subsubsection{Local UV field}

Our model shows that the OH mid-IR lines trace the photodissociation of H$_2$O by photons in the range $114-143~$nm. As such, mid-IR OH lines carry information on the local UV radiation field in this range. In the following, we note $F_{UV}^{\tilde{B}}$, the photon flux integrated between $114$ and $143$~nm, which is $F_{UV}^{\tilde{B}}= 6\times 10^{7}$~cm$^{-2}$~s$^{-1}$ for a Draine radiation field. The photodissociation rate $k_{\tilde{B}}$ is the integral over the wavelength of the local UV radiation field multiplied by the photodissociation cross section:
\begin{equation}
k_{\tilde{B}} = \int_{114~\rm{nm}}^{143~\rm{nm}} I(\lambda) \sigma(\lambda) d\lambda.
\label{eq:k-sigma}
\end{equation}
Strictly speaking, the conversion between $k_{\tilde{B}}$ and the local UV flux $F_{UV}^{\tilde{B}}$ should then rely on the knowledge of the shape of the UV radiation field. However, for the considered radiation fields, the intensity weighted cross section averaged between $114$ and $143$~nm varies only by a factor of $\sim 3$ and is about $\sigma \simeq 6 \times 10^{-18}$~cm$^{2}$. Thus, Eq. (\ref{eq:k-sigma}) can be approximated by
\begin{equation}
F_{UV}^{\tilde{B}} \simeq  \frac{k_{\tilde{B}}}{6 \times 10^{-18}~\rm{cm}^{2}},
\label{eq:measurement-UVFlux}
\end{equation}
with a typical uncertainly of about a factor of three.


\subsection{Application to HH~211}

\begin{figure*}
\centering
\includegraphics[width=1.\textwidth]{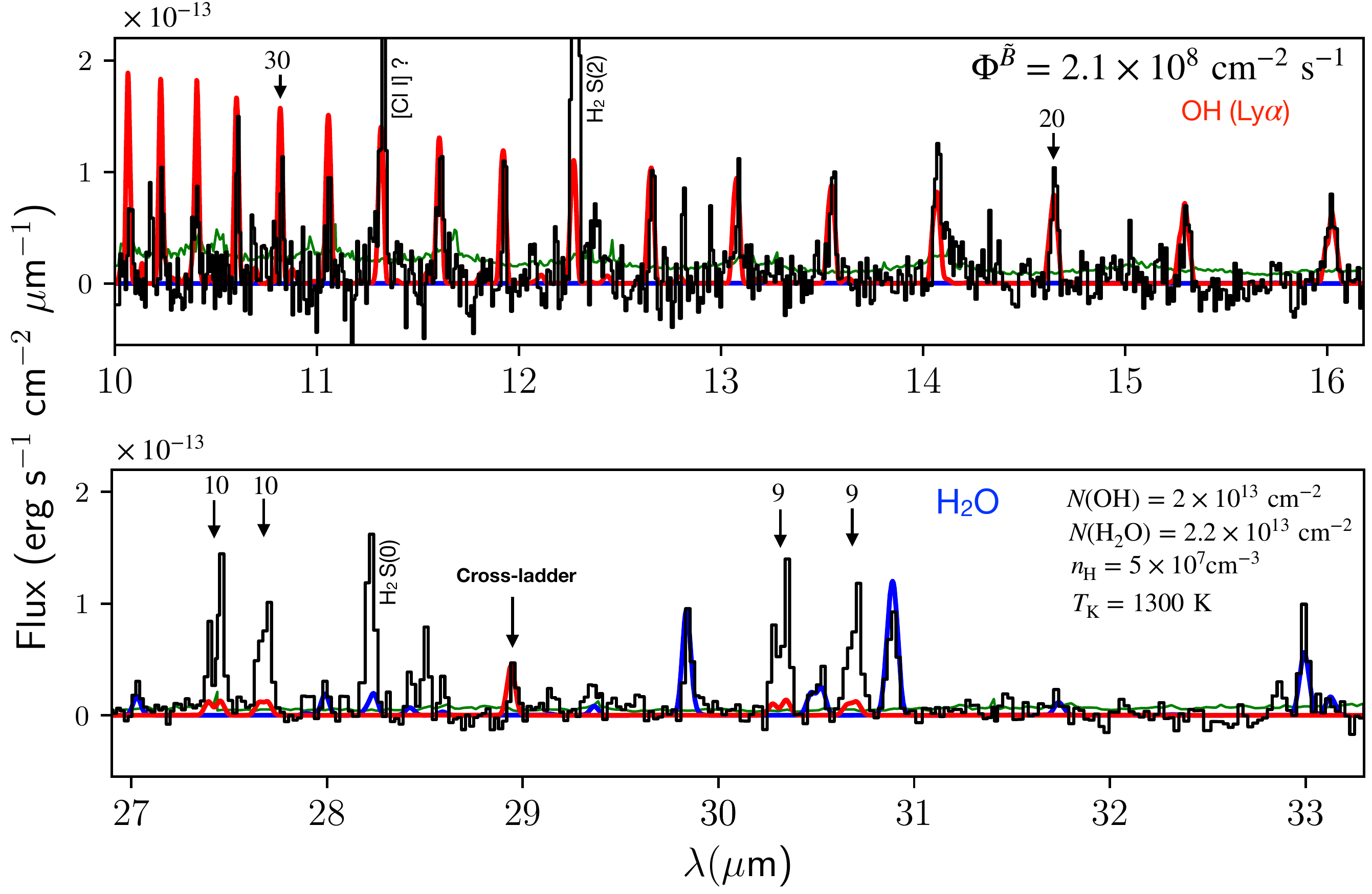}
\caption{Comparison of the \textit{Spitzer}-IRS mid-IR spectrum of the tip of HH 211 outflow from \citet{2008ApJ...680L.117T} (black lines) and of a synthetic {\tt GROSBETA} spectrum for OH (in red) and H$_2$O (in blue). The best fit parameters are indicated in the top right parts of the panels. \revbis{The {\tt GROSBETA} spectra are plotted assuming a source size equal to the extraction aperture $\Delta \Omega = 1.9\times 10^{-9}$ sr.} The rotational quantum number $N_{\text{up}}$ are indicated for some OH lines. The model reproduces well the OH lines at 10-16~$\mu$m, but underproduces them at longer wavelengths.}
\label{fig:HH211-vs-grosbeta}
\end{figure*}

The \textit{Spitzer}-IRS observations of the apex of the young protostellar jet HH 211 provide one of the best examples of highly rotationally excited OH. Figure \ref{fig:HH211-vs-grosbeta} shows the unique sequence of superthermal OH emission lines unveiled from $10~\mu$m down to $30~\mu$m coming from rotational levels between $N=34$ to $N=9$. The association between OH mid-IR emission and ongoing H$_2$O photodissociation is further supported by the detection of a compact H$\alpha$ emission by \citet{2006AJ....132..467W}, tracing a strong UV emitting shock in the close vicinity of the OH emission. Complementary observations of H$_2$ rovibrational emission \citep[2.12~$\mu$m, ][]{2006ApJ...636L.141H} and in the sub-millimeter domain unveiled a complex structure of multiple molecular bow-shocks encompassing the strong shock \citep{2012ApJ...751....9T}.

\subsubsection{Diagnostics from $N>20$ lines}
\label{subsubsec:chemical-pumping}
\rev{From the \textit{Spitzer}-IRS spectrum published by \citet{2008ApJ...680L.117T}, we measure a flux of the OH line at 14.6~$\mu$m of $F = 3.5 \times 10^{-15}$ erg~s$^{-1}$~cm$^{-2}$ in an aperture of $\Omega = 1.9\times 10^{-9}$~sr.} This line emerges from the $N=20$ level of the OH($X$)($\varv=0$) state and corresponds to an intensity integrated over solid angle of \rev{$I = 2.3\times 10^{-5}$~erg~s$^{-1}$~cm$^{-2}$} \revbis{(see Eq. (\ref{eq:flux-intensity}))}. Figure \ref{fig:measuring-FUV-145nm} indicates that the conversion factor for each component of the quadruplet is $\sim 0.2$, regardless the spectral shape of the radiation field. At \textit{Spitzer}-IRS spectral resolution ($R \simeq 600$ for the short–high module), the $\Lambda$-doublet and the fine-structure is however not spectrally resolved. Equation (\ref{eq:measurment-phi}) then yields a column density of H$_2$O photodissociated per unit time of
\begin{equation}
\Phi^{\tilde{B}} =  \frac{I}{4 \times 0.2 h \nu}  =  \rev{2.1 \times 10^{8}}~ \rm{cm}^{-2}~\rm{s}^{-1},
\label{eq:phi-HH212}
\end{equation}
where the factor 4 stands for the sum over the four components of quadruplet, \rev{and $\nu$ is the frequency of the line}. This corresponds to a total number of H$_2$O molecule photodissociated per second of $4 \times 10^{41}$~molecule~s$^{-1}$.

Figure \ref{fig:HH211-vs-grosbeta} (top panel) compares a synthetic spectrum of OH produced by a Ly$\alpha$ radiation field with the \textit{Spitzer}-IRS spectrum between 10 and 16~$\mu$m. We recall that in this wavelength range, the spectrum depends only on $\Phi^{\tilde{B}}$ and not on \nH, \TK~or \TIR. Our model reproduces well the sequence of the lines coming from $N=31$ down to $N=14$. 

The higher three lines lying in the range 10-10.5$~\mu$m are overestimated by the model. The extinction correction \citep[$A_{\rm{v}} = 10$~mag,][]{2006AJ....132..467W} cannot explain this discrepancy since the extinction curve used by \citet{2008ApJ...680L.117T} tends to overestimate the extinction at 10~$\mu$m. 
\LE{An ISRF UV radiation field, which produces relatively weaker line fluxes shortward of 11~$\mu$m does not significantly improve the fit.} Alternatively, we posit that the rotational state of the parent H$_2$O could modify the rotational distribution above $N \simeq 30$. JWST observations with a higher signal-to-noise ratio, combined with modeling including the rotational state of the parent H$_2$O are required to clarify this.

$\Phi^{\tilde{B}}$ can then be used to derive $k_{\tilde{B}}$, provided the column density of H$_2$O is known. Rotational lines of H$_2$O are detected from the far-IR with \textit{Herschel}-PACS to the mid-IR with \textit{Spitzer}-IRS, spanning energy levels from 110 up to 1800~K. The mid-IR lines have been analyzed using {\tt GROSBETA} and we derive a column density of \rev{$2.2^{+2.0}_{-1.2} \times 10^{13}$~cm$^{2}$}, a density of \nH$\simeq \rev{5 \times 10^7}$~cm$^{-3}$, and a temperature of \TK$\simeq 1300~$K (see synthetic spectrum in Fig. \ref{fig:HH211-vs-grosbeta}, bottom panel). Interestingly, the same model reproduces the excitation temperature derived from the observations of far-IR lines \citep[$T_{ex} \simeq 90$~K,][]{2018A&A...616A..84D}.  
This suggests that both mid-IR and far-IR lines trace the same warm H$_2$O reservoir that is photodissociated. 
Assuming a common origin for the OH and H$_2$O lines, Eq. (\ref{eq:measurement-kb}) yields
\begin{equation}
\rev{k_{\tilde{B}} \simeq 9.5\pm4 \times 10^{-6}~\rm{s}^{-1}.}
\label{eq:k-HH211}
\end{equation}

The strong shock revealed by compact H$\alpha$ emission is expected to produce a UV field dominated by Ly$\alpha$ photons. Equation (\ref{eq:k-sigma}) then gives \rev{$F_{UV}^{\tilde{B}} \simeq 8.5 \times 10^{11}$~photon~cm$^{-2}$~s$^{-1}$}, which corresponds to a UV photon flux \rev{$5 \times 10^{3}$} times larger than that of the Draine ISRF in the range $91.2-200$~nm. 
\rev{This value is well in line with models of fast dissociative shocks with $V_S \simeq 50$~km~s$^{-1}$ and \nH $\simeq 3 \times 10^{5}$cm$^{-3}$ \citep[][]{1979ApJS...39....1R,2020A&A...643A.101L}, assuming that H$_2$O is produced in the warm postshock without further dilution of the UV radiation field. An origin of OH and H$_2$O emission in slower shocks passively illuminated by the stronger dissociative shock is also possible \citep[see models of][]{2015ApJ...806..227M,2019A&A...622A.100G}. Spatially resolved observations with JWST-MIRI of both OH mid-IR lines and atomic lines that trace fast dissociative shocks such as [NeII] at $12.8 \mu m$ or [Ni II] at $6.6 \mu m$ \citep{1989ApJ...342..306H} are required to distinguish the two possible scenarios.}

\subsubsection{Evidence of additional chemical pumping routes}
Figure \ref{fig:HH211-vs-grosbeta}-b shows that the OH lines coming from lower $N$ levels are detected with the \textit{Spitzer}-IRS long-high module. In particular, the cross-ladder transition at 28.9 $\mu m$ is well reproduced by our {\tt GROSBETA} model with a column density of \rev{$N($OH$) = 2 \times 10^{13}$cm$^{-2}$, assuming a density of \nH$=5 \times 10^{7}$cm$^{-3} $ and a temperature of \TK$=1300$~K derived from the H$_2$O lines. This model is illustrative since the emission of the OH cross-ladder line might originates from a different layer of gas but its suggests a $N($OH$)/N($H$_2$O$)$ ratio of about $\simeq 0.9$. This value lies in the upper part of the range $10^{-3}$ to $0.8$ typically derived toward protostellar outflows, and is higher than those derived toward warm inner envelopes \citep{2011A&A...531L..16W,2012A&A...548A..77G,2015ApJ...799..102G}.  \revbis{Interestingly, our derived ratio of $\simeq 0.9$ is closer} to that estimated in low $A_{\rm{V}}$ regions of the Orion Bar \citep[$\gtrsim 1$,][]{2011A&A...530L..16G}. This further supports the driving role of UV photodissociation in preventing the full conversion of OH into H$_2$O and in maintaining a relatively high $N($OH$)/N($H$_2$O$)$ ratio under warm conditions. Interestingly, detailed  models of molecular shocks do predict that both, fast dissociative shocks and passively irradiated slow shocks exhibit high $N($OH$)/N($H$_2$O$)$ ratios \citep{1989ApJ...340..869N,2019A&A...622A.100G,2020A&A...643A.101L}.}

For this set of {\tt GROSBETA} parameters, the intra-ladder rotational lines longward of $\lambda = 18~\mu$m are underestimated by our {\tt GROSBETA} model by up to a factor of eight. As shown in Sec. \ref{subsubsec:transition-mid-far}, these intermediate-$N$ lines ($9<N<15$) are excited either by prompt emission, or by \rev{IR radiative pumping or collisions}. In HH~211, the IR background detected by \textit{Spitzer}-IRS is too weak to have a significant impact on the excitation of these levels. In fact, for the selected set of parameters, the excitation is dominated by prompt emission (right part of the schematic diagram, Fig. \ref{fig:OH-spectra-transition}-c). 
One could invoke collisional excitation to take over from prompt emission and increase the intensities of the intra-ladder transitions. This can be achieved by either decreasing the chemical pumping rate $\mathcal{D}$ or by increasing \nH~and/or \TK~(i.e., increasing $\mathcal{D}_{crit}$, see  Fig. \ref{fig:OH-spectra-transition}-c). Because $\Phi = \mathcal{D} N($OH$)$ is constrained by the lines in the range 10 to 16$~\mu$m, the intensity of the chemical pumping rate $\mathcal{D}$ can only be decreased by increasing $N($OH$)$. However, increasing $N($OH$)$, \nH~or \TK~would increase the intensities of the cross-ladder transitions, which are already well reproduced by our model. In other words, collisions or IR radiative pumping cannot reconcile our models with observations. 

Interestingly, in this wavelength range, $\Lambda$ doublets are spectrally resolved and show an asymmetry between $A'$ and $A''$ $\Lambda$-doublet states of about a factor of 1.5-2. As discussed by \citet{2014ApJ...788...66C}, this further supports the minimal impact of collisional (de)excitation since collisions might eliminate asymmetry in the population. This suggests that the intermediate-$N$ transitions reveal another excitation process that is not included in our model.

The reaction H$_2$+O can produce OH in rotationally and vibrationally excited states and supplement the excitation of intermediate-$N$ levels. In particular, experimental data show that the reaction
\begin{equation}
\rm{H}_2 + O(^1 D) \rightarrow OH + H
\label{eq:H2OD-OH}
\end{equation}
produces OH with rotational number between $N =10$ and $25$ with a large fraction produced in $\varv \ge 1$ and a propensity to the $A'$ states \citep{2000Sci...289.1536L}. Interestingly, the discrepancy between our predictions and the \textit{Spitzer}-IRS spectra increases from $N \sim 16$ down to $N \sim 9$. The case of chemical excitation by reaction (\ref{eq:H2OD-OH}) can be treated in a similar way as we described the excitation of OH following H$_2$O photodissociation (see Sec. \ref{subsec:mid-IR}). In the case of H$_2$O photodissociation, the line intensities are given by the column density of OH formed by H$_2$O photodissociation. Similarly, reaction (\ref{eq:H2OD-OH}) will increase the line intensities in proportion to the number of OH produced per second via this reaction. The discrepancy between the model and the observations suggests that the formation rate of OH via Eq. (\ref{eq:H2OD-OH}) is about 6 times larger than that via H$_2$O photodissociation. Interestingly, at the density derived by our non-LTE analysis of H$_2$O lines, O($^1$D) would be converted into OH instead of decaying via the radiative transition at  $\lambda=$6300$~\AA$. However, this scenario requires an efficient production of O($^1$D), either by OH or H$_2$O photodissociation, or by collisional excitation. \BT{Still, the estimated column density ratio OH/H$_2$O appears to be too low for OH photodissociation to take over from H$_2$O photodissociation in the excitation of OH mid-IR lines. Moreover, at Ly$\alpha$, only} $\sim 10\%$ \BT{of the water photodissociation ends in O($^1$D) \citep{1982JChPh..77.2432S,2008JPCA..112.3002V}, a fraction that is too small for the required efficiency of O($^1$D) production.} Alternatively, the reaction between O($^3$P) and vibrationally excited H$_2$, known to be present at the tip of the bow-shock, could lead to OH in rotationally (and vibrationally) excited states \citep{2011ApJ...735...90G, 2014ApJ...788...66C}. \rev{These additional chemical pumping routes could also impact the excitation of even lower-$N$ lines.} Further modeling combining OH excitation and chemistry is required to clarify the excitation of the intermediate-$N$ lines.

\subsection{Perspectives with JWST-MIRI}

\begin{figure}
\centering
\includegraphics[width=.48\textwidth]{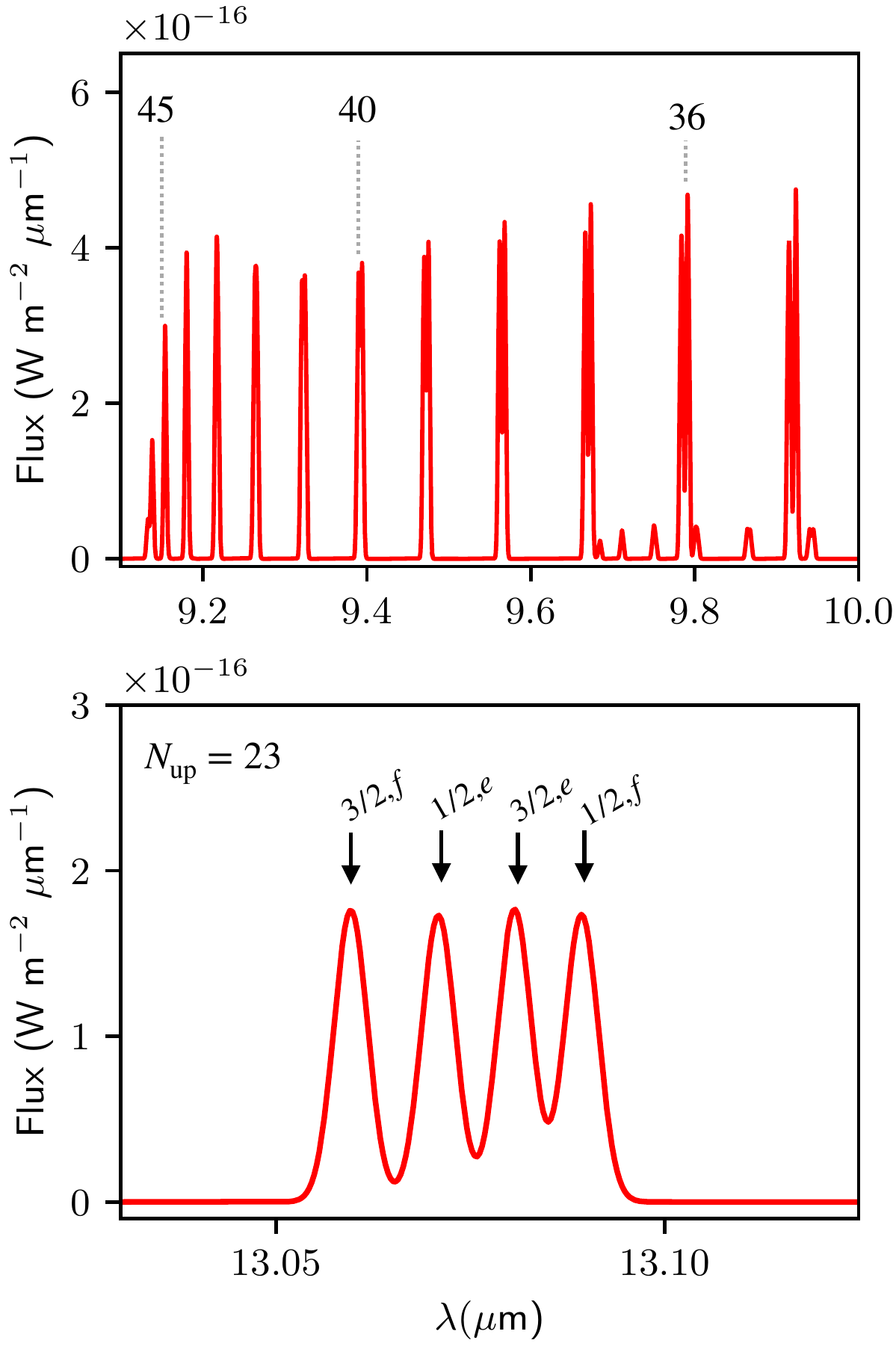}
\caption{\rev{Expected OH mid-IR spectrum at JWST-MIRI} resolving power ($R\simeq 2000-3000$). The parameters of the model correspond to those derived from the \textit{Spitzer}-IRS observations of HH 211. The weak lines around 9.8~$\mu$m are pure rotational lines within the OH($X$)($\varv=1$) state.  
}
\label{fig:MIRI}
\end{figure}

JWST-MIRI will provide a unique view in the mid-IR regime by combining spectroscopic and imaging capabilities \citep{2015PASP..127..584R,2015PASP..127..595W}. At the observed wavelength range ($5-28~\mu$m), JWST will be able to detect pure rotational lines coming from $N \ge 10$ and cross-ladder lines coming from $N \ge 5$, allowing for detailed studies of the excitation of OH across the full rotational ladder. In this section, we focus on the high-$N$ rotational lines. 

Up to now, OH mid-IR lines tracing H$_2$O photodissociation have been detected toward three protostellar outflows \citep{2012ApJ...751....9T} and a large number of protoplanetary disks \citep{2010ApJ...720..887P,2011ApJ...731..130S,2011ApJ...733..102C,2014ApJ...788...66C}. We predict that the jump in sensitivity 
will extend the detectability of high-$N$ lines to other irradiated environments. Since the line intensities depend on the local abundance of H$_2$O and on the local radiation field, only astrochemical models designed for specific sources can robustly predict mid-IR line intensities. However, the detectability of the mid-IR lines as a function of the physical conditions can be explored with simple chemical arguments. The line emissivity per hydrogen atom depends on $x($H$_2$O$) G_0$, where $x($H$_2$O$)$ is the H$_2$O abundance with respect to the total density of H atoms and $G_0$ the local UV flux (see Eq. (\ref{eq:1D})). \rev{In warm and irradiated environments (fast shocks, disk surface layers, low $A_{\rm{V}}$ depths of prototypical PDRs), one can assume that H$_2$O is destroyed by photodissociation and formed in the gas phase by two-body reactions so that $x($H$_2$O$) \propto$~\nH$/G_0$}. \rev{For cooler regions, for which H$_2$O vapor is produced by photodesorption or by gas-phase ion-neutral routes, the scaling is expected to be different. Therefore, at least for warm environments,} the emissivity of the OH lines per total hydrogen is expected to scale with the density. This may explain why superthermal OH emission has been detected by \textit{Spitzer} in dense environments and remains undetected is classical photodissociation regions \citep{2011A&A...530L..16G}. We thus posit that high-$N$ lines could be detected by JWST-MIRI in more diffuse regions. Observation of dense classical PDRs such as the Orion bar or NGC 2023 should be considered as primary targets. In even more diffuse environments, such as translucent clouds, we expect OH mid-IR lines to remain undetected where we estimate $\Phi \simeq 10^{3}-10^{4}$~cm$^{-2}$~s$^{-1}$ \citep{2013ApJ...762...11F}, \rev{corresponding to line intensities as low as $\sim 10^{-12}-10^{-11}$~erg~s$^{-1}$~cm$^{-2}$~sr$^{-1}$.}

The sensitivity of JWST will also allow us to probe the mid-IR emission in deeply embedded protostars. At high extinction ($A_{\rm{V}} \gtrsim 20$), the OH lines in the range 13-15~$\mu$m will be particularly suited to minimize the IR extinction. The OH mid-IR emission can then constitute a unique tool to unravel the poorly known physical and chemical structure of young protostellar disks. Regarding protostellar jets and outflows, it will allow us to test if the low water abundance derived from \textit{Herschel} observations is due to UV photodissociation or other physical processes \citep{2014A&A...572A...9K}.

The spatial resolution provided by mapping capabilities will also be crucial to constrain the emitting region of OH and map the photodissociation of H$_2$O. For dense environments, maps of $N($H$_2$O$)$ can be built from pure rotational lines of H$_2$O lying typically longward of $20~\mu$m ($E_{\rm{up}} \lesssim 1200~$K) or ro-vibrational lines and used to construct maps of the local UV radiation field (see Eq. (\ref{eq:measurement-UVFlux})). 

The Medium Resolution Spectrometer (MIRI-MRS), with its relatively high spectral resolution ($R \sim 2000-3000$) will provide a valuable astrophysical view of H$_2$O photodissociation that will challenge current quantum calculations. Figure \ref{fig:MIRI} shows a synthetic spectrum at MIRI-MRS spectral resolution. First, MIRI-MRS will give access to the OH lines in the range 9-10$~\mu$m that are directly populated by H$_2$O photodissociation ($35<N_{\rm{up}}<46$, see also Fig. \ref{fig:distrib-RF}). The $\Lambda$-doublets up to $N_{\rm{up}} = 40$ and the fine-structure up to $N_{\rm{up}} = 29$ will be spectrally resolved. This information will help to further constrain the shape of the UV radiation field and understand the impact of the rotational state of the parent H$_2$O. New quantum calculations, including the rotational state of H$_2$O, and the fine-structure and the $\Lambda$-doubling of the OH product, are warranted.

%
%
%
%

%

\section{Conclusion}
\label{sec:concl}

In this work, we explore the potential of the OH($X$)($\varv=0,N$) mid-IR emission to probe H$_2$O photodissociation. To reach this goal, results from quantum mechanical calculations resolving the electronic, vibrational and rotational state of the OH product following H$_2$O photodissociation at different UV wavelengths are collected. The distribution of the OH photofragments is then calculated for UV fields of various spectral shapes and included in a new radiative transfer code called {\tt GROSBETA}. The impact of prompt emission (i.e., formation of OH in excited states following H$_2$O photodissociation), collisional excitation and radiative pumping is extensively studied.

The main conclusions of this study are:
\begin{enumerate}

\item The mid-IR emission of OH in the range 9-16~$\mu$m is an unambiguous tracer of H$_2$O photodissociation. The mid-IR emission is the result of the rotational radiative cascade of OH photofragments within the $\varv=0$ state. As such, the detection of these lines constitutes a prime evidence for the presence of active H$_2$O chemistry. In particular, the line intensities are directly proportional to the column density of H$_2$O photodissociated per second by photon in the range $114-143$~nm, denoted as $\Phi^{\tilde{B}}$. The conversion factors between the line intensities and $\Phi^{\tilde{B}}$ is provided for each rotationally excited line and for UV radiation fields of various shapes in Fig. \ref{fig:measuring-FUV-145nm} and in Appendix \ref{app:conversion}. These conversion factors depend little on the exact spectral shape of the UV radiation field.
\item Provided a measurement of the column density of the irradiated water is known, the photodissociation rate by photons in the range $114-143$~nm can be inferred (see Eq. (\ref{eq:measurement-kb})). The UV flux can then be deduced with good accuracy, regardless of the shape of the UV radiation field. Alternatively, provided an estimate of the strength of the local UV radiation field is available, the column density of H$_2$O can be derived.
\item The precise state distribution of the OH fragments depends on the spectral shape of the UV radiation field and results in small differences in the shape of mid-IR spectrum of OH. This suggests that the relative line intensities, if accurately measured, can be used to derive constraints on the spectral energy distribution of the UV radiation field.
\item The lower rotational levels, probed either by cross-ladder rotational lines shortward of $30~\mu$m, or by far-IR rotational lines, are excited by IR radiative pumping or collisions. We provide criteria to determine if a rotational level of an intermediate energy (typically $E_{\text{up}} =3000$ K$, N \simeq 10 $) is excited by prompt emission or thermal processes.
\item As a test case, our model is applied to an irradiated region located at the tip of the HH 211 protostellar jet. Our simple single-zone model reproduces the \textit{Spitzer}-IRS spectrum in the range 11-16$~\mu$m and shows that water is exposed to a UV photon flux that is about $\sim 5 \times 10^3$ times larger than the standard UV interstellar radiation field. Discrepancies between our predictions and the observations at longer wavelength suggest that additional chemical pumping processes such as the reaction H$_2$+O might contribute to the excitation of the rotational levels between $N\sim 10-20$.
\item JWST-MIRI will probe the rotational ladder of OH from $N = 5 $ up to $N \simeq 45$ corresponding to energy levels from $800~$K up to $45 000$~K. We posit that the jump in sensitivity will unveil OH mid-IR emission in new environments such as classical photodissociation regions (e.g., Orion bar) and embedded protostars. Moreover, armed with good spectral resolution ($R\sim 2000-3000$), the MIRI-MRS observing mode will give access to even more excited lines in the  range $9-10~\mu$m and to the fine-structure and $\Lambda$-doubling state distribution of the OH fragments. 
\end{enumerate}

\LE{Our work demonstrates the potential of OH mid-IR lines to probe H$_2$O photodissocation and the strength of the UV field. New quantum calculations and astrochemical models taking into account the quantum state of the species are warranted to leverage future JWST observations and unveil oxygen chemistry through the excitation of OH.}

\begin{acknowledgements}
\rev{\revbis{The authors thank the anonymous referee for their constructive comments.} B.T. would also like to thank B. Godard, A. Bosman and A. Faure for fruitful discussions.}
This work is part of the research programme Dutch Astrochemistry Network II with project number 614.001.751, which is (partly) financed by the Dutch Research Council (NWO).
\end{acknowledgements}

\bibliographystyle{aa} 
\bibliography{export-bibtex.bib} 

\begin{appendix}


%
%
%

\section{Distribution of OH fragments}
\label{app:distrib}


\begin{figure}[!t]
\centering
\includegraphics[width=0.49\textwidth]{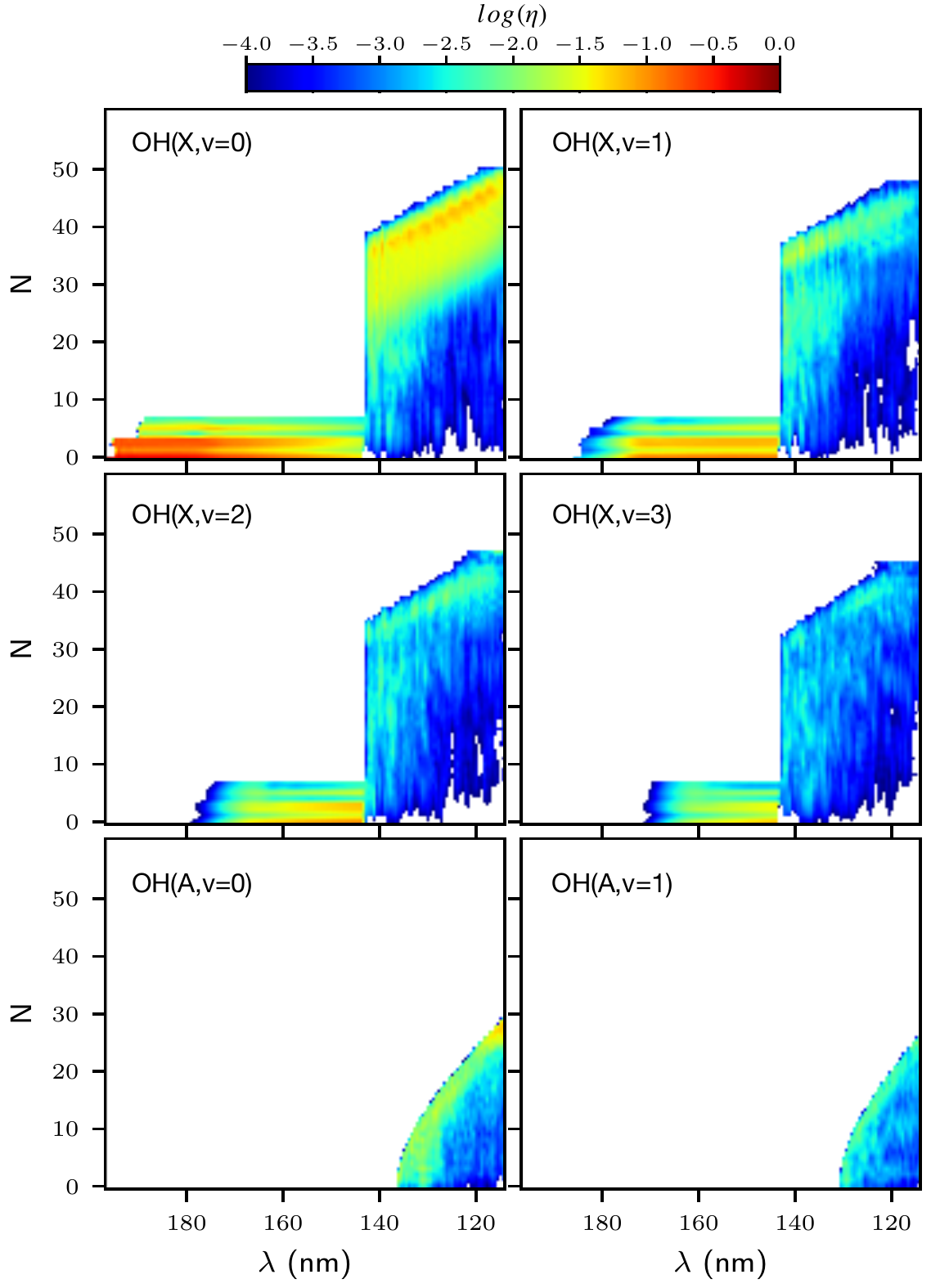} 
\caption{Rotational distributions of nascent OH for various vibrational and electronic states as a function of the photodissociation wavelength. The OH distributions are summed over fine-structure and $\Lambda$-doubling states. The difference in the rotational distribution between photodissociation through the $\tilde{A}$ ($\lambda > 143~$nm) and $\tilde{B}$ ($\lambda < 143~$nm) states of H$_2$O is clearly visible.}
\label{app:eta}
\end{figure}

\begin{figure}
\centering
\includegraphics[width=0.48\textwidth]{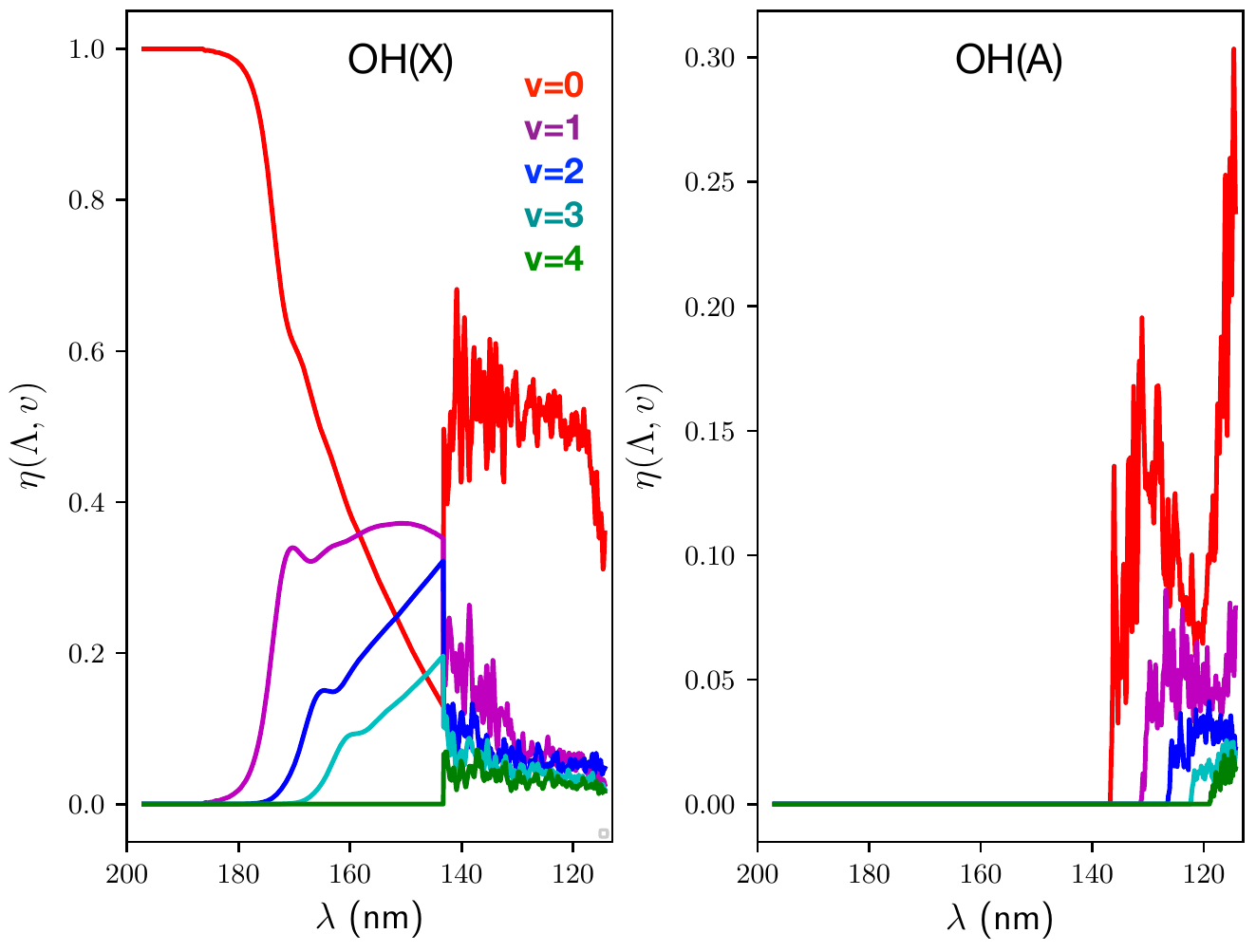} 
\caption{Vibrational distributions of nascent OH for the two electronic states of OH as a function of the UV wavelength. The distributions $\eta$ are summed over rotational, fine-structure and $\Lambda$-doubling states. The difference in the vibrational distribution between photodissociation through the $\tilde{A}$ ($\lambda > 143~$nm) and $\tilde{B}$ ($\lambda < 143~$nm) states of H$_2$O is clearly visible.}
\label{app:eta-vib}
\end{figure}

\BT{In this appendix, we present the state distribution of OH following H$_2$O photodissociation as a function of the UV wavelength implemented in our model. The data stem from the quantum calculations published in \citet{2000JChPh.112.5787V} and \citet{2001JChPh.114.9453V}. For the latter, the calculations have been repeated using an another potential energy surface that gives better agreement with experiments (see Sec. \ref{subsec:chem-dataset}). As discussed in Sec. \ref{subsec:chem-dataset}, we denote as $\eta(\lambda,\Lambda, \varv,N)$ the probability to form OH in the state OH($\Lambda$)($\varv, N$) following H$_2$O photodissociation by a photon of wavelength $\lambda$. Figure \ref{app:eta} shows the rotational distributions of OH as function of the photon wavelength for different electronic and vibrational states.
Figure \ref{app:eta-vib} gives the vibrational distributions for each electronic state. The distributions shown in Fig. \ref{app:eta} and \ref{app:eta-vib} do not account for the subsequent dissociation of the OH products. For example, the OH($A$) rotational levels with $\varv \ge 2$ are dissociative and are thus discarded in our excitation calculations.}

\begin{figure}
\centering
\includegraphics[width=0.5\textwidth]{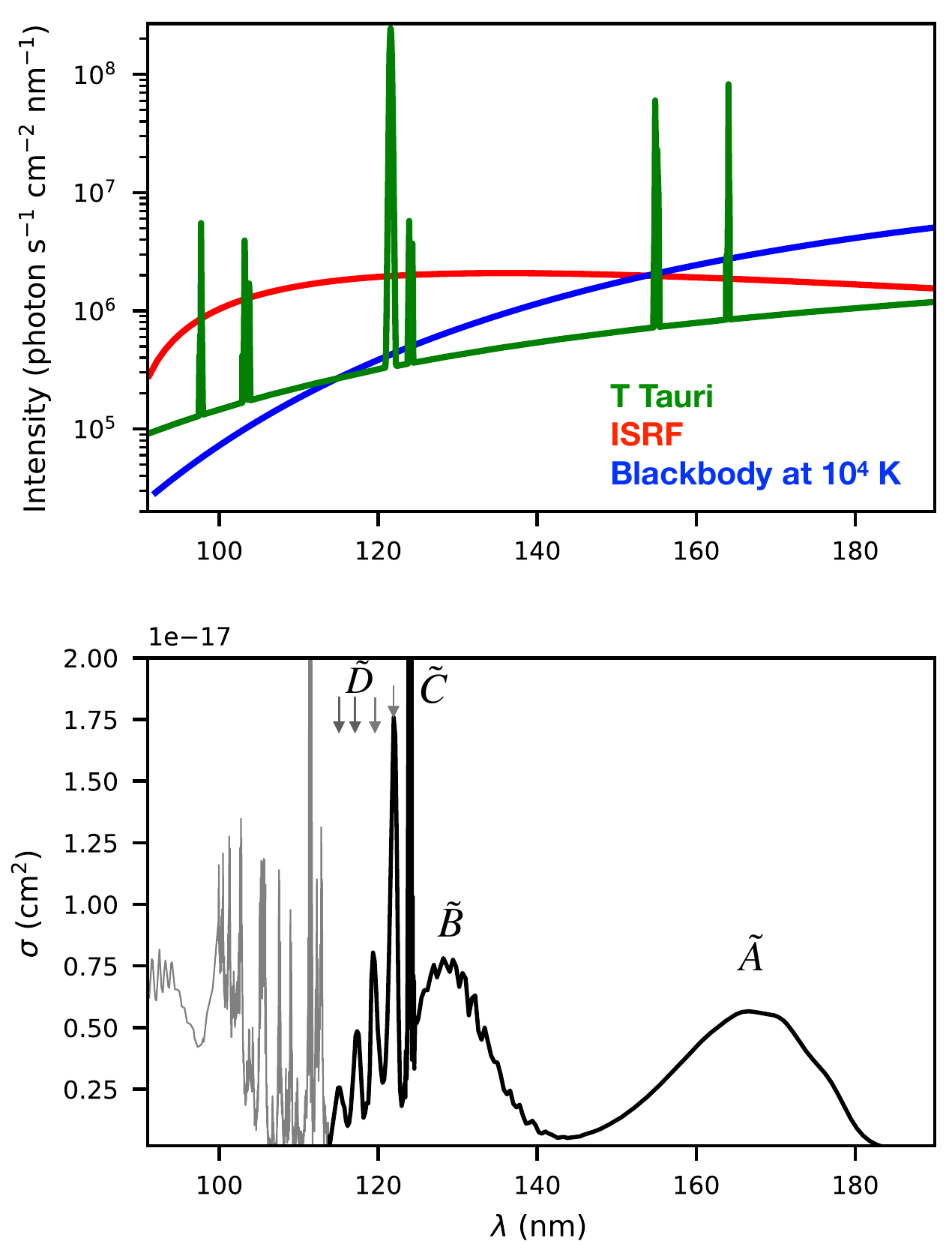}
\caption{Radiation fields and H$_2$O photodissociation cross section for photodissociation adopted in this work. \textit{Top:} Spectral shape of the radiation fields as a function of wavelength. The radiation fields have been scaled to agree with the integrated energy-intensity of the Draine 1978 radiation field between 91.2 and 200~nm, that is $2.6 \times 10^{-6}$~W~m$^{-2}$. \textit{Bottom:} Photodissociation cross section of H$_2$O producing OH adopted in this work (black) and the extension to 911 \AA~(gray). The features due to photodissociation through the $\tilde{A}$, $\tilde{B}$, $\tilde{C}$, and $\tilde{D}$ states of H$_2$O are also indicated.}
\label{app:cross-section-H2O}
\end{figure}

\BT{In this work, the state distribution of the OH fragments following H$_2$O photodissociation by UV radiation fields of different shapes are computed using Eq. (\ref{eq:def-ki}) and (\ref{eq:def-fi}). Figure \ref{app:cross-section-H2O} presents the spectral shape of the radiation fields adopted in this work. The radiation fields representative of an accreting T Tauri star and of the interstellar radiation field corresponds to those used by \citet{2017A&A...602A.105H} and are available at \url{https://home.strw.leidenuniv.nl/~ewine/photo/}. The UV radiation field named "Ly$\alpha$" is a single Ly$\alpha$ line ($\lambda= 121.567$~nm) with a Doppler broadening parameter of $b=200$ km s$^{-1}$.} \BT{The total photodissociation cross section of H$_2$O leading to OH (Fig. \ref{app:cross-section-H2O}) is collected by \citet{2017A&A...602A.105H} from \citet{2003JPhB...36.2767F,2004JChPh.120.6531F}, \citet{2005CPL...416..152M}, and \citet{2008JPCA..112.3002V}.}

\section{Collisional rate coefficients}
\label{app:collisional-rates-coeff}

The collisional rate coefficients between OH and H$_2$ computed by \citet{1994JChPh.100..362O} that include levels up to $N=5$ have been extrapolated to higher $N$ numbers assuming that the downward rates follow
\begin{equation}
k_{N \rightarrow N'} = g_{N'} a e^{-b \Delta E/T_K},
\end{equation}
where $N$ and $N'$ designate the rotational quantum number of the upper and lower levels, $\Delta E$ the energy difference between these levels and $g_{N'}$ the degeneracy of the lower energy level. The parameters $a$ and $b$ are the best fit coefficients of available data at temperature $T_K$. Collisions with the ortho and para states of H$_2$ have been considered separately. In this work, $T_K \ge 500~$K and we adopted the rate coefficient computed at a maximum kinetic temperature of $T_K =300~$K by \citet{1994JChPh.100..362O} without any further extrapolation. \BT{The relative proportion of ortho and para populations of H$_2$ is set to 3.} We estimate that the extrapolated rate coefficients are accurate within a factor of ten. Collisional (de)excitation involving vibrational and electronic states is not considered here.
\revbis{Collisional excitation by electrons and H can also be relevant in certain irradiated environments. However, to our knowledge, the corresponding rate coefficients have not yet been computed: Thus, we neglected the contribution of electrons and H to the excitation of OH.}
\newpage

\section{Analytical model for mid-IR lines}

\label{app:anal-model}

\begin{figure}
\centering
\includegraphics[width=.46\textwidth]{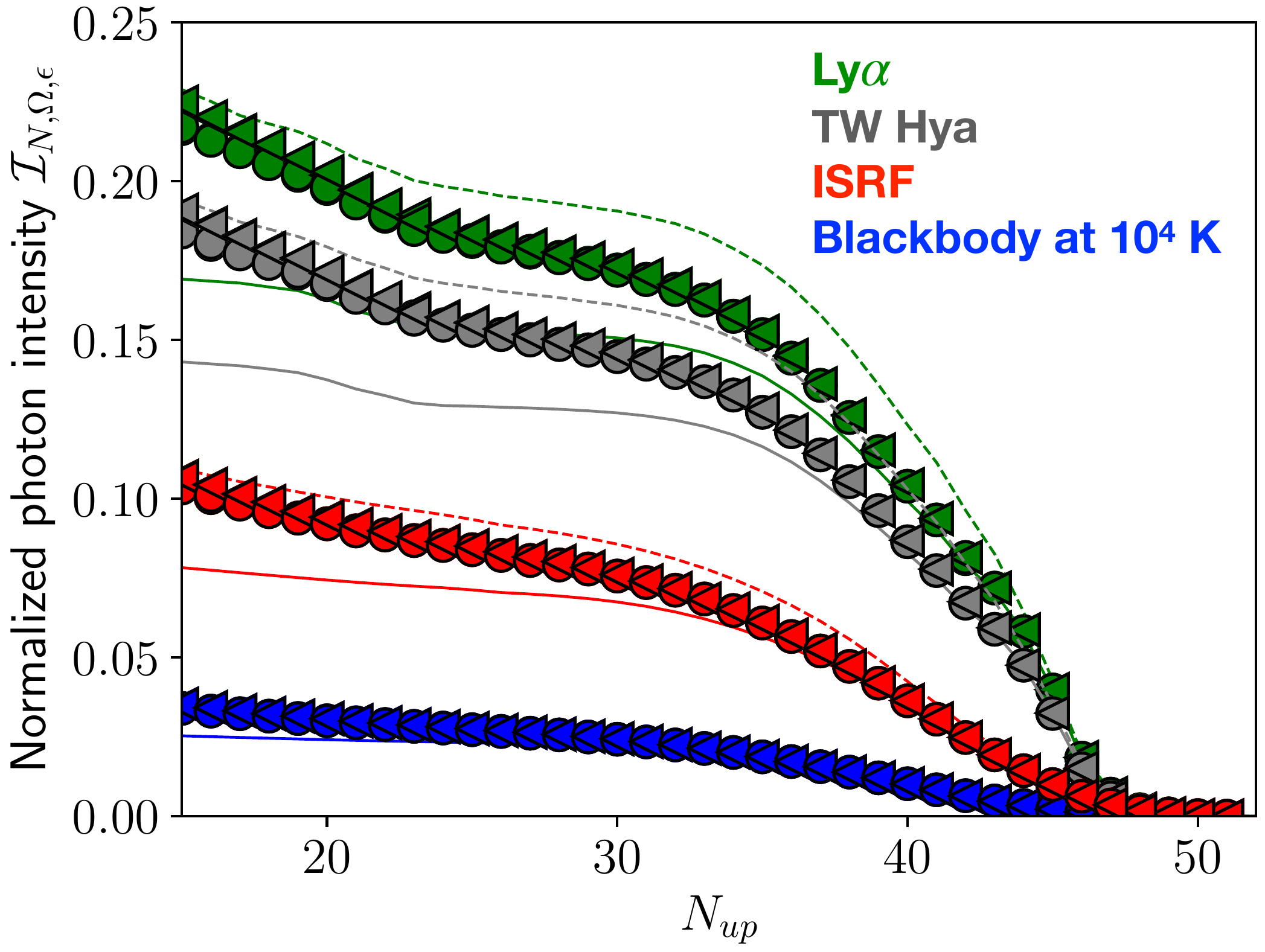}
\caption{Normalized photon intensity of the OH($X^2\Pi$)($\varv=0$) intra-ladder mid-IR lines as a function of the rotational number of the upper energy level. The intensities are computed for the fiducial parameters given in Table \ref{table:param} and for three spectral shapes of the UV field (color coded), and divided by $\Phi$ according to Eq. \ref{eq:normalised-int}. Circle and triangle markers correspond to lines belonging to the $\Omega = 1/2$ and $\Omega = 3/2$ ladders, respectively. $\Lambda$-doublets are indiscernible in this plot. In this regime, rotational levels are only populated by the radiative cascade of OH photofragments. The solid line is the analytical model assuming that the intra-ladder lines within the $\varv=0$ state are produced by the radiative cascade of OH fragments produced in the $\varv=0$ state (Eq. (\ref{eq:high-N-intensity})). The dotted line is the analytical model that assumes that any OH produced in a vibrational state instantaneously decays toward the ground vibrational state with a negligible change of its rotational number (see Eq. (\ref{eq:FN-model-B})).}
\label{app:measuring-FUV-145nm}
\end{figure}

The link between the state distribution of OH fragments $f_i$ (Fig. \ref{fig:example-distrib} and \ref{fig:distrib-RF}) and the variation of the line intensities with the upper $N $ number (see Fig. \ref{fig:lines-lya-mid-IR} and \ref{fig:lines-grid-mid-IR}), quantified here by $\mathcal{I}_N$, can be clarified by a simple analytical model of the radiative cascade. Neglecting collisional (de)exciatation and radiative pumping, and assuming that the fraction of OH in rotationally excited states constitutes a negligible fraction of the total population of OH, the statistical equilibrium equation (see Eq. (\ref{eq:statistical-eq-red})) for high-$N$ states yields
\begin{equation}
\sum_{i>j} M_{ij} N_{j} + \Phi \times  f_i  = 0~~\text{with}~~ M_{ij}=\begin{cases}
     A_{ij}  & (E_i > E_j)\\
     - A_{ij} & (E_i < E_j),
  \end{cases}
\label{eq:radiative-cascade-1}
\end{equation}
where $N_{i}$ is the column density of OH in the state $i$, and $A_{ij}$ the Einstein-A coefficient of the spontaneous emission $i\rightarrow j$.
This equation being linear in $N_i$ and the \rev{right hand side} term depending linearly on $\Phi$, shows that $N_i$ is proportional to $\Phi$ and that the relative population $N_i / N(\rm{OH})$ depends only on the Einstein-$A$ coefficients and on $f_i$ but not on $\Phi$. Consequently, the linear relation between high-$N$ line integrated intensities and $\Phi$ shown in Fig. \ref{fig:lines-lya} (top right panel), as well as the fact that the global shape of the mid-infrared spectrum is independent of $\Phi$, is a characteristic feature of the radiative cascade.

The radiative matrix $M_{ij}$ in Eq. (\ref{eq:radiative-cascade-1}) can be greatly reduced by noting that the radiative decays of the $N$ levels within the $X ^{2}\Pi(\varv=0)$ state are dominated by pure rotational $N\rightarrow N-1$ transitions. Moreover, excited electronic states are found to rapidly decay to the ground electronic state with little change of their vibrational and rotational quantum number. By neglecting the contribution of vibrationally excited levels to the populations of OH($\varv=0, N$) and assuming that all OH photofragments produced in the OH(A$^2\Sigma^+$) ($\varv =0,N$) state decay immediately to the OH($X^2\Pi$) ($\varv =0,N$) state, Eq. (\ref{eq:radiative-cascade-1}) applied to the OH($X$)($\varv=0,N$) levels yields

\begin{equation}
A_{N+1 \rightarrow N} N_{N+1} - A_{N \rightarrow N-1} N_{N} + \Phi \times \tilde{f}(N, \varv=0)  \simeq 0,
\label{eq:radative-cascade-2}
\end{equation}
with  

\begin{equation}
\tilde{f}(N, \varv=0) = f(X,N,\varv=0) + \frac{1}{2} f(A,N,\varv=0),
\label{eq:tilde-f}
\end{equation}
where the factor $\frac{1}{2}$ stands for the different degeneracies between the OH($X$)($\varv, N$) and OH($A$)($\varv, N$) states. The solution of Eq. (\ref{eq:radative-cascade-2}) is then

\begin{equation}
A_{N \rightarrow N-1} N_{N} \simeq \tilde{I}_N \Phi,
\label{eq:radative-cascade-3}
\end{equation}
where
\begin{equation}
\tilde{I}_N \simeq \sum_{N' \ge N} \tilde{I}(N', \varv=0)
\label{eq:FN-model-A}
\end{equation}
is the probability to form OH in the ground vibrational state with a rotational number larger than $N$. The high-$N$ lines are found to be optically thin. Their integrated intensities are then given by
\begin{equation}
I_1(N \rightarrow N-1) =  \tilde{I}_N \frac{h c}{\lambda} \Phi.
\label{eq:high-N-intensity}
\end{equation}
Thus, according to our analytical model, the normalized integrated photon intensity $\mathcal{I}_{N}$ corresponds to $\tilde{I}_{N}$, the probability to form an OH photofragment in a $\varv =0$ state with a rotational number $N'$ larger than $N$. In particular, the line intensities $N \rightarrow N-1$ do not depend on the Einstein-$A$ coefficients. This equation reflects the fact that any OH produced in an excited rotational level $N'$ will eventually decay through the $N \rightarrow N-1$ transition with $N \le N'$. Because each rotational state of OH is split into two spin-orbit substates that are further subdivided into two $\Lambda$-doubling states, $\tilde{I}_N \le 1/4$, as seen in Figs. \ref{fig:lines-lya-mid-IR} and \ref{app:measuring-FUV-145nm}.  


As discussed in Sec. \ref{subsec:mid-IR} and further shown in Fig. \ref{app:measuring-FUV-145nm}, our model underestimates the line intensities computed with {\tt GROSBETA}. This indicates that the radiative decay from vibrationaly excited levels contributes to the population of $X^2\Pi (N, \varv=0)$ levels. In order to take into account the contribution of vibrationaly excited levels, we assume that any OH produced in a vibrationally excited state immediately decays toward the  $X^2\Pi (\varv=0)$ state with no change of the rotational number. This leads to a normalized integrated photon intensity of
\begin{equation}
\tilde{I}_N \simeq  \sum_{\varv \ge 0} \sum_{N' \ge N} \tilde{I}(N', \varv).
\label{eq:FN-model-B}
\end{equation}
This model always overestimates the line intensities (dashed lines, Figs. \ref{fig:lines-lya-mid-IR} and \ref{app:measuring-FUV-145nm}). This is due to the fact that for $N \gtrsim 15$, Einstein-$A$ coefficients of rovibrational transitions are smaller than those of pure rotational transitions. Thus, an OH fragment produced in a $\varv > 0$ state tends to decay through the intra-ladder transition within its nascent vibrational state before decaying to the $\varv=0$ state with a lower $N$ number than its nascent one. Below $N \simeq 25$, the analytical model reproduces well the computed intensities as all levels produced in higher $N$ numbers in $\varv >0$ have decayed toward the $\varv = 0$ states. The lines emerging from $N < 25$ are thus tracing the production of OH in high rotational states, even though the decay of electronically excited OH in the ground vibrational state also contributes to the variation of $\mathcal{I}_N$ below $N \simeq 25$.

\section{Conversion factors}
\label{app:conversion}
\BT{Table \ref{table:conversion-factors} gives the line intensities of the OH mid-IR lines in the pure radiative cascade regime normalized by $\Phi^{\tilde{B}}$ denoted as $\mathcal{I}^{\tilde{B}}$ (see Eq. (\ref{eq:normalised-int-B})). The intensity of the $N_{up} \rightarrow N_{up}-1$ lines of the OH($X$)($\varv=0$) state are summed over the four components of the quadruplet. $\mathcal{I}^{\tilde{B}}$ can be used to convert any line intensity into a column density of H$_2$O photodissociated per second through the H$_2$O $\tilde{B}$ state denoted as $\Phi^{\tilde{B}}$ using Eq. (\ref{eq:measurment-phi}).}

\begin{table}
\caption{Conversion factors between the photon intensity integrated over solid angle (in photon cm$^{-2}$ s$^{-1}$) of the rotational lines $N_{up} \rightarrow N_{up}-1$ summed over the components of each quadruplet and $\Phi^{\tilde{B}}$ (molecule cm$^{-2}$ s$^{-1}$).}              
\centering                                      
\begin{tabular}{c c c c c}          
\hline\hline                        
$N_{up}$  & Ly$\alpha$ & T Tauri & ISRF & BB $10^4$~K \\    
\hline                                   
 10 &    0.96 &    0.94 &    0.88 &    0.88 \\
 11 &    0.95 &    0.92 &    0.86 &    0.85 \\
 12 &    0.93 &    0.91 &    0.84 &    0.83 \\
 13 &    0.92 &    0.89 &    0.82 &    0.81 \\
 14 &    0.90 &    0.88 &    0.80 &    0.79 \\
 15 &    0.88 &    0.86 &    0.78 &    0.77 \\
 16 &    0.87 &    0.84 &    0.76 &    0.75 \\
 17 &    0.85 &    0.83 &    0.74 &    0.73 \\
 18 &    0.83 &    0.81 &    0.72 &    0.71 \\
 19 &    0.82 &    0.80 &    0.70 &    0.69 \\
 20 &    0.80 &    0.78 &    0.69 &    0.68 \\
 21 &    0.78 &    0.76 &    0.67 &    0.66 \\
 22 &    0.76 &    0.74 &    0.66 &    0.65 \\
 23 &    0.75 &    0.73 &    0.65 &    0.63 \\
 24 &    0.74 &    0.72 &    0.64 &    0.62 \\
 25 &    0.73 &    0.71 &    0.63 &    0.61 \\
 26 &    0.72 &    0.70 &    0.61 &    0.60 \\
 27 &    0.72 &    0.69 &    0.60 &    0.58 \\
 28 &    0.71 &    0.69 &    0.59 &    0.57 \\
 29 &    0.70 &    0.68 &    0.58 &    0.56 \\
 30 &    0.69 &    0.67 &    0.56 &    0.54 \\
 31 &    0.68 &    0.66 &    0.55 &    0.52 \\
 32 &    0.67 &    0.64 &    0.53 &    0.50 \\
 33 &    0.65 &    0.63 &    0.51 &    0.48 \\
 34 &    0.63 &    0.61 &    0.48 &    0.45 \\
 35 &    0.61 &    0.59 &    0.45 &    0.42 \\
 36 &    0.58 &    0.56 &    0.42 &    0.38 \\
 37 &    0.54 &    0.52 &    0.39 &    0.35 \\
 38 &    0.50 &    0.49 &    0.35 &    0.31 \\
 39 &    0.46 &    0.44 &    0.31 &    0.27 \\
 40 &    0.42 &    0.40 &    0.27 &    0.23 \\
 41 &    0.37 &    0.36 &    0.23 &    0.19 \\
 42 &    0.33 &    0.31 &    0.18 &    0.14 \\
 43 &    0.29 &    0.27 &    0.14 &    0.11 \\
 44 &    0.23 &    0.22 &    0.11 &    0.08 \\
 45 &    0.16 &    0.15 &    0.07 &    0.05 \\
 46 &    0.07 &    0.07 &    0.05 &    0.03 \\
 47 &    0.03 &    0.03 &    0.03 &    0.02 \\
 48 &    0.01 &    0.01 &    0.01 &    0.01 \\
 49 &    0.00 &    0.00 &    0.01 &    0.00 \\
\hline   
\end{tabular}
\label{table:conversion-factors}
\end{table}

\section{Critical pumping rate}
\label{app:crictical-pumping}

\begin{figure}[!t]
\centering
\includegraphics[width=.49\textwidth]{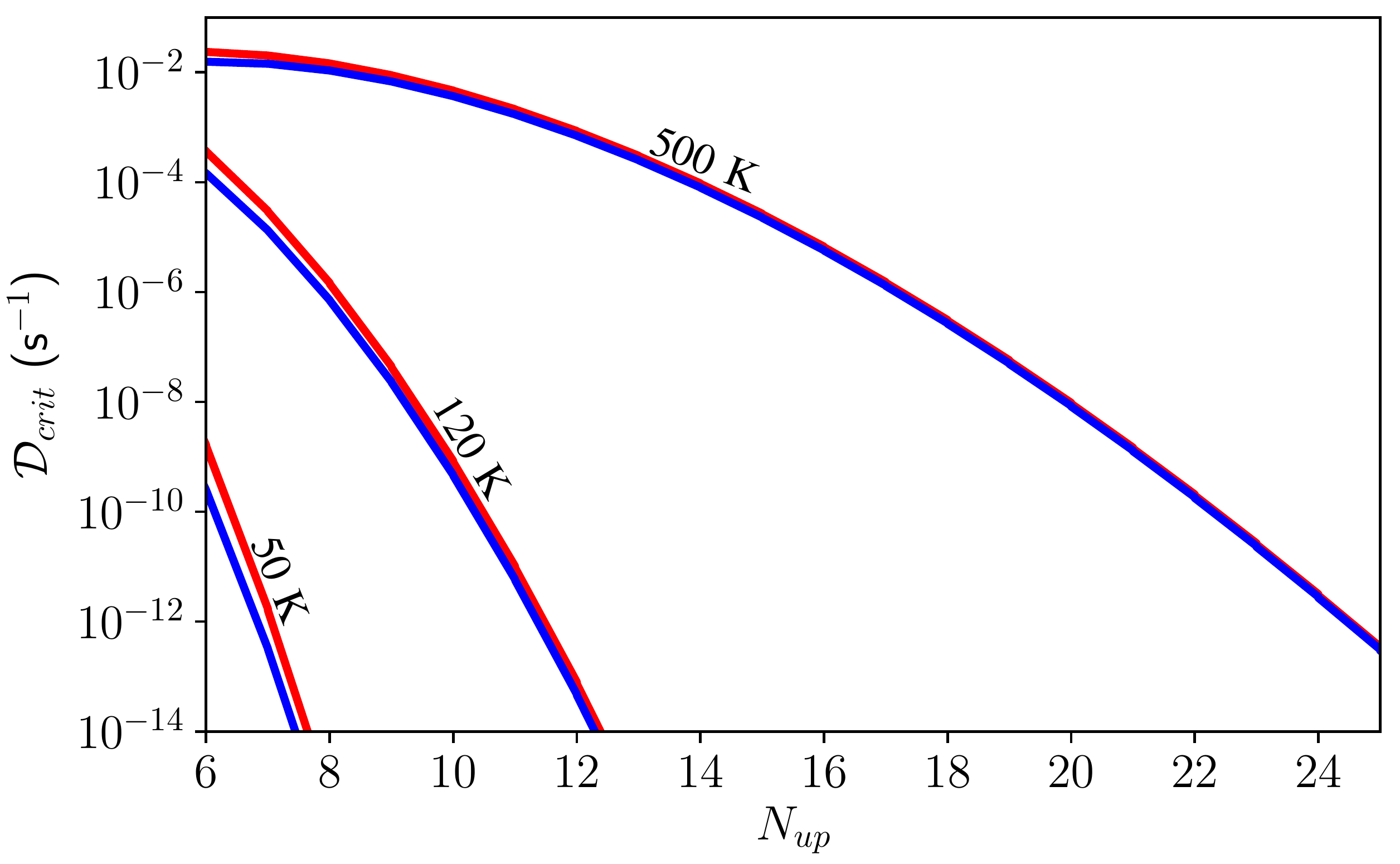} 
\caption{Critical value of the chemical pumping rate $\mathcal{D}_{\rm{crit}} \equiv \left( \Phi/N(\text{OH}) \right)_{\rm{crit}} $ below which populations in the $X^2\Pi(\varv=0,N,\Omega)$ states are populated by collisions or IR radiative pumping processes rather than by prompt emission. \rev{Collisions or IR radiative pumping} are assumed to populate the levels according to a Boltzmann distribution of a single temperature $T_{\rm{ex}}$. Because Einstein-$A$ coefficients differ between the two rotational ladders, two $\mathcal{D}_{\rm{crit}}$ exist for each $N$ quantum number. For simplicity, we have assumed $\mathcal{I}_N = 0.25$ for all $N$ (see sec. \ref{app:anal-model}).}
\label{fig:D-crit}
\end{figure}
Figure \ref{fig:OH-spectra-transition}-c summarizes the dominant excitation processes as a function of the density~\nH~ and the chemical pumping rate $\mathcal{D} = \Phi/N(\rm{OH})$ for a given OH line. The parameter space is split in two regions: For low values of $\mathcal{D}$, the excitation is dominated thermal \rev{and radiative} processes whereas for high values of $\mathcal{D}$ the excitation is set by the radiative cascade of the OH photofragments.

The boundary between the two regions is defined by $\mathcal{D}_{crit}$ (green line Fig. \ref{fig:OH-spectra-transition}-c). Its value depends on the physical conditions of the gas, namely \nH, \TIR, and \TK. For example, Fig. \ref{fig:OH-spectra-transition}-a shows that $\mathcal{D}_{crit}$ increases from $2 \times 10^{-9}$ to $2 \times 10^{-8}$~s$^{-1}$ by increasing the density from \nH=$10^{7}$ to $10^{9}$\cmsq. We propose here to derive simple estimates of $\mathcal{D}_{\rm{crit}}$ for any rotational level and any \TK~and \TIR. To do so, we compare the intensity predicted by a pure radiative cascade (see Eq. (\ref{eq:normalised-int}) and Fig. \ref{fig:lines-lya-mid-IR}) to that produced by collisions and/or IR radiative pumping only. In the pure IR radiative pumping regime (regime \ding{173}), the line intensity is given by Eq. (\ref{eq:low-N-intensity}). This yields to a critical chemical pumping rate above which prompt emission sets the population of a level $N$ of
\begin{equation}
\mathcal{D}_{\rm{crit}}(T_{\rm{ex}}, N) = \frac{g_{N} A_{N \rightarrow N-1}}{\mathcal{I}_N Q(T_{\rm{ex}})} e^{-E_{N}/k_B T_{\rm{ex}}},
\label{eq:critical-rate}
\end{equation}
where $T_{\rm{ex}}=$\TIR. This equation can also be directly transposed to the high density regime for which $T_{\rm{ex}} \simeq T_{\rm{K}}$ (regime \ding{175}). For intermediate densities, for which both collisions and IR radiative pumping are relevant (regime \ding{174}), $\mathcal{D}_{\rm{crit}}$ varies smoothly from $\mathcal{D}_{\rm{crit}}(T_{\rm{IR}})$ to $\mathcal{D}_{\rm{crit}}(T_{\rm{K}})$ (see Fig. \ref{fig:OH-spectra-transition}-c).

Figure \ref{fig:D-crit} shows $\mathcal{D}_{\rm{crit}}$ as a function of the upper $N$ number of the level and as a function of the excitation temperature. $\mathcal{D}_{\rm{crit}}(T_{\rm{ex}}, N)$ decreases with $N$, demonstrating that lines with higher $N$ are less sensitive to thermal or radiative excitation processes and more sensitive to prompt emission. This is mostly due to the fact that these lines are coming from levels that are much higher in energy than \TK~or \TIR. For higher temperatures, thermal processes are more efficient at populating higher $N$ levels resulting in an increase in $\mathcal{D}_{\rm{crit}}$. We also note that the decrease in $\mathcal{D}_{\rm{crit}}$ with $N$ is stiff, showing that for given chemical pumping rate, the transition between lines excited by prompt emission and lines excited by IR radiative pumping or collisions is well defined. In particular, we recover the fact that when collision are negligible and for \TIR$=120~$K, $\mathcal{D}_{crit} \simeq 10^{-9}$~s$^{-1}$ for the line coming from $N = 10$. In contrast, a chemical pumping rate of at least $10^{-4}$~s$^{-1}$ is required to excite the line coming from $N = 6$. For high-$N$ levels prompt emission dominates for rates as low as $\mathcal{D} \simeq 10^{-12}$~s$^{-1}$. This shows that the schematic view of the parameter space proposed in Fig. \ref{fig:OH-spectra-transition}-c remains valid for the low-$N$ lines for which the boundary is shifted to the right, and for high-$N$ lines, for which the boundary is shifted to the left by orders of magnitudes.

\end{appendix}

\end{document}